# Constructing Recursion Operators in Intuitionistic Type Theory

Lawrence C Paulson
Computer Laboratory
Corn Exchange Street
Cambridge CB2 3QG
England

October 1984

## Abstract

Martin-Löf's Intuitionistic Theory of Types is becoming popular for formal reasoning about computer programs. To handle recursion schemes other than primitive recursion, a theory of well-founded relations is presented. Using primitive recursion over higher types, induction and recursion are formally derived for a large class of well-founded relations. Included are < on natural numbers, and relations formed by inverse images, addition, multiplication, and exponentiation of other relations. The constructions are given in full detail to allow their use in theorem provers for Type Theory, such as Nuprl. The theory is compared with work in the field of ordinal recursion over higher types.

# Contents

# 1   Introduction

Per Martin-Löf originally developed his Intuitionistic Type Theory [15] as a formal notation for constructive mathematics, but it may also be regarded as a programming language [14]. It supports the principle of *propositions as types*: there is a type forming operator for each logical connective and quantifier, with typing rules that correspond to the intuitionistic rules of natural deduction. This type structure is rich enough to completely specify computational problems: Nordström and Smith specify a program to produce a KWIC index [17]. Its potential for the systematic development for correct programs is attracting the interest of computer scientists [1, 3].

Many people have used the interactive theorem prover LCF to verify functional programs by computer. Most LCF work until now [10, 18] has been conducted in a logic for domain theory. Petersson [21] has already implemented a version of LCF for Intuitionistic Type Theory (henceforth Type Theory). Constable's group has implemented an elaborate theorem prover, called Nuprl, for a version of Type Theory [6]. I am also experimenting with Type Theory in a new theorem prover, Isabelle [19].

Type Theory has no explicit principle of general recursion: it forces all functions to terminate by allowing only primitive recursion. Many functions that are not primitive recursive can be defined as primitive recursive *functionals*: the classic example is Ackermann's function [16]. The study of functionals leads to a simple theory of types, including at least the natural numbers **Nat**, and function types $A \to B$ if $A$ and $B$ are types [27]. Each type $A$ can be assigned a level $L(A)$ such that $L(\mathbf{Nat}) = 0$ and $L(A \to B)$ is the greater of $L(A) + 1$ and $L(B)$. (This scheme of type levels is too simple for Intuitionistic Type Theory.) Primitive recursive functionals are also called primitive recursive functions of *higher type*.

Primitive recursive functionals do not consititute a natural programming language. To facilitate reasoning about programs in Type Theory, this paper formalizes a more conventional programming style: writing arbitrary recursion equations annotated with a termination proof.

Every computer scientist knows of Quicksort, a total function not obviously definable by primitive recursion. Jan Smith [24] defines Quicksort in Type Theory by first deriving appropriate rules of induction and recursion, using primitive recursion over higher types. The "while rule" of Backhouse and Khamiss [1] is also a form of primitive recursion. This paper considers induction and recursion over a wide class of well-founded (w.f.) relations.

The rule of induction over a well-founded relation $\prec$ is

$$\frac{\forall x.\,(\forall x' \prec x.\,P(x')) \to P(x)}{\forall x.\,P(x)}\ .$$

Classically, this rule is sound if $\prec$ has no infinite descending chains $x_1 \succ x_2 \succ \cdots$. W.f. induction (or Noetherian induction) is often used in program verification, and is the fundamental method for proving termination [11]. Total functions can be defined by the related principle of w.f. recursion.

Even if it is apparent that a relation is w.f., proving this may be difficult. W.f. relations are most easily constructed from simpler ones, using rules that preserve the w.f. property. Later sections present Type Theory derivations of induction and recursion over

- the ordering $<$ on the *natural numbers*
- a *subrelation* of a w.f. relation
- the *inverse image* of a w.f. relation
- the *transitive closure* of a w.f. relation
- the *disjoint sum* of two w.f. relations
- the *lexicographic product* of two w.f. relations
- the *lexicographic power* of a w.f. relation
- the *immediate subtree* relation on a wellordering type.

These ideas have been extensively studied by both computer scientists and logicians. Manna and Waldinger [13] describe and verify similar rules for w.f. relations. The sole induction principle of the Boyer-Moore theorem prover [2] is w.f. induction, over lexicographic products of inverse images of $<$. The theory of *ordinal recursion* concerns recursion over total wellorderings constructed by addition, multiplication, and exponentiation of simpler wellorderings [26]. These operations suffice to reach wellorderings of order type $\epsilon_0$, which is the first ordinal $\alpha$ to satisfy $\omega^\alpha = \alpha$. The functions definable by ordinal recursion up to $\epsilon_0$ include all functions provably recursive in Peano arithmetic. Terlouw's work on ordinal recursion over higher types [27] is related to the rule for powers in section 12.

It should not surprise anyone that various principles of w.f. recursion can be developed within Intuitionistic Type Theory. The point of this paper is to present the derivations in *full* detail. They can be used with a mechanical theorem prover like Nuprl. The present set of w.f. relations is complete in two senses. (1) It includes nearly all the w.f. relations I have encountered in the literature. (2) The set includes order types well beyond $\epsilon_0$, indeed any ordinal that a proof of program termination could reasonably require.

The paper assumes a basic knowledge with Type Theory. Good introductions include Martin-Löf [14, 15] and Nordström and Smith [17]. Jan Smith [25] presents the inference rules and gives a formal interpretation of the semantics. Each section begins with an easy introduction, then becomes increasingly formal. At any point you may skip to the next section. The remaining sections

- describe a notation for Type Theory;
- describe the LCF style of backwards proof used in this paper;

- formalize w.f. relations, induction, and recursion operators in Type Theory;

- define the function Quicksort using a recursion operator, and show that it satisfies the usual recursion equations;

- derive w.f. induction and recursion for each rule for constructing w.f. relations,

- show that w.f. relations are precisely the inverse images of wellordering types;

- use the rules to construct w.f. relations taken from the literature;

- comment on drawbacks and questions about the approach.

## 2 Notation

There is a lamentable diversity of notations for Type Theory. Constable and Backhouse have completely different notations; Martin-Löf's has evolved between his earlier and later papers. The notation of Chalmers University [17] is Martin-Löf's except for the names of the selectors. The application of the function $f$ to arguments $a_1, \ldots, a_n$ is written $f(a_1, \ldots, a_n)$; the abstraction of an expression $c$ over the variables $x_1, \ldots, x_n$ is written $(x_1, \ldots, x_n)c$. A function of several arguments is regarded as a function-valued function, a device known as *currying*; so $f(a, b)$ abbreviates $f(a)(b)$, and $(x, y)c$ abbreviates $(x)(y)c$.

My notation for Type Theory is similar. Types include

$\bot$, the empty type

$\top$, the type containing the one value 0

**Bool**, the type of truth values $\{\mathbf{T}, \mathbf{F}\}$

$a =_A a'$, the equality type $\mathbf{Eq}(A, a, a')$, where $A$ is a type (the subscript $A$ may be omitted if obvious from context)

The selectors take their arguments in an unconventional order: the argument $p$ being eliminated appears last rather than first. If the constructions $c, c_0, c_1, \ldots$ have type $C$, then so do the following selections. Each is listed together with Martin-Löf's version [15]:

For $p \in \bot$, I use **contr**$(p)$ instead of $R_0(p)$.

For $p \in$ **Bool**, I use **cond**$(c_1, c_2, p)$ instead of $R_2(p, c_1, c_2)$.

For $p \in A \times B$, I use **split**$((x, y)c, p)$ instead of $E(p, (x, y)c)$.

For $p \in A + B$, I use **when**$((x)c_1, (y)c_2, p)$ instead of $D(p, (x)c_1, (y)c_2)$.

For $p \in$ **Nat**, I use **natrec**$(c_0, (x, y)c_1, p)$ instead of $R(p, c_0, (x, y)c_1)$, similarly for **listrec**, **transrec**, etc.

This argument order works well with curried functions. For instance, **cond**$(c_1, c_2)$ means the same as $(p)$**cond**$(c_1, c_2, p)$; it is a function on type **Bool**, mapping **T** to $c_1$ and **F** to $c_2$. If $f$, $g$, and $h$ are appropriate functions, then **split**$(f)$ is a function with domain $A \times B$ and **when**$(g, h)$ is a function with domain $A + B$. The advantages of this argument order are apparent in complex expressions involving selectors, such as

$$\lambda(\mathbf{when}(\mathbf{natrec}(p,q),\mathbf{split}((x)\,\mathbf{split}((y,z)r(x,y,z)))))\ ,$$

which could have type $(\mathbf{Nat}+(A\times(B\times C)))\to D$. The standard argument order requires additional variables $v_1, v_2, v_3, v_4$, denoting intermediate values that we have no wish to see:

$$\lambda v_1.\,\mathbf{when}(v_1,(v_2)\,\mathbf{natrec}(v_2,p,q),(v_3)\,\mathbf{split}(v_3,(x,v_4)\,\mathbf{split}(v_4,(y,z)r(x,y,z))))$$

Actually, the details of selectors are not important; in my presentation of proof constructions, most selectors are banished in favor of new functions and equations defining them.

The pair of $a$ and $b$ is written $\langle a,b\rangle$. The functions **fst** and **snd** are special cases of **split**, with $\mathbf{fst}(\langle a,b\rangle)=a$ and $\mathbf{snd}(\langle a,b\rangle)=b$. The application of the function object $f$ to the object $a$ is written $f*a$ rather than the usual $\mathbf{apply}(f,a)$. The canonical objects of a function type have the form $\lambda((x)b)$, which may equivalently be written in the traditional lambda style, $\lambda x.b$. When $d$ is a large expression, $\lambda(d)$ is simpler and clearer than $\lambda x.d(x)$.

Here are a few inference rules in the notation:

$\to$ and $\Pi$ elimination

$$\frac{f\ \in\ \prod_{x\in A} B(x) \qquad a\in A}{f*a\in B(a)}$$

$\times$ and $\Sigma$ introduction

$$\frac{a\in A \qquad b\in B(a)}{\langle a,b\rangle\ \in\ \sum_{x\in A} B(x)}$$

$\times$ and $\Sigma$ elimination

$$\frac{\begin{array}{c}[x\in A;\ y\in B(x)]\\ c(x,y)\in C(\langle x,y\rangle)\end{array}\qquad p\ \in\ \sum_{x\in A} B}{\mathbf{split}(c,p)\ \in\ C(p)}$$

$+$ elimination

$$\frac{\begin{array}{cc}[x\in A] & [y\in B]\\ c_A(x)\in C(\mathbf{inl}(x)) & c_B(y)\in C(\mathbf{inr}(y))\end{array}\qquad p\in A+B}{\mathbf{when}(c_A,c_B,p)\ \in\ C(p)}$$

**Nat** elimination

$$\frac{c_0\in C(0) \qquad \begin{array}{c}[x\in\mathbf{Nat};\ u\in C(x)]\\ c_1(x,u)\in C(\mathbf{succ}(x))\end{array} \qquad p\in\mathbf{Nat}}{\mathbf{natrec}(c_0,c_1,p)\ \in\ C(p)}$$

Classically, a wellordering is a linear well-founded relation. Martin-Löf's *wellordering type* is a general kind of tree structure. Both uses of 'wellordering' are firmly established. To prevent confusion, I always write 'wellordering type' for Martin-Löf's usage, and 'wellordering' for the classical usage.

The rule for wellordering elimination uses the latest innovation, hypothetical hypotheses. A hypothesis of the form $f(x)\in B\ [x\in A]$ asserts that $f$ is a function from $A$ to $B$.

This is hardly different from the assumption $f \in A \to B$, but allows the wellordering types to be defined without mention of function types. The types in Type Theory are defined independently of each another.

$$\cfrac{p \in \mathbf{W}_{x \in A} B(x) \qquad \begin{bmatrix} x \in A \\ f(z) \in \mathbf{W}_{x \in A} B(x) \quad [z \in B(x)] \\ u(z) \in C(f(z)) \quad [z \in B(x)] \\ c(x, f, u) \in C(\mathbf{sup}(x, f)) \end{bmatrix}}{\mathbf{transrec}(c, p) \in C(p)}$$

I make occasional, non-essential use of the *subtype* $\{x \in A \mid B(x)\}$. This is similar to the type $\sum_{x \in A} B(x)$, but its elements are those of $A$ instead of pairs $\langle a, b \rangle$. Subtypes are convenient when $B(x)$ represents a proposition such as equality, whose elements are uninteresting. Constable gives a fuller explanation [5, page 74], as does Petersson [22]. Rules include

type formation

$$\cfrac{A \text{ type} \qquad \begin{array}{c}[x \in A] \\ B(x) \text{ type}\end{array}}{\{x \in A \mid B(x)\} \text{ type}}$$

introduction

$$\cfrac{a \in A \quad b \in B(a)}{a \in \{x \in A \mid B(x)\}}$$

elimination

$$\cfrac{a \in \{x \in A \mid B(x)\} \qquad \begin{array}{c}[x \in A;\ B(x) \text{ true}] \\ c(x) \in C(x)\end{array}}{c(a) \in C(a)}$$

The assumption $B(x)$ **true** means that $B(x)$ is a true proposition, namely that the type $B(x)$ contains an element. Such reasoning can be formalized by introducing a set of rules about true propositions. An alternative is to regard this assumption as an abbreviation of $y \in B(x)$, with the restriction that $y$ must not appear free elsewhere in the rule.

## 3  Backwards proof

A traditional method of searching for a proof is to work backwards from goals to subgoals. For instance, to prove $A \wedge B$, it suffices to prove $A$ and $B$ separately. Backwards proof is more difficult in Type Theory: the goal would be written $p \in A \times B$, for an *unknown* construction $p$. As the proof proceeds, constraints upon $p$ accumulate; when the proof is finished, they determine $p$ completely. Recording the constraints requires tedious bookkeeping, whether the proof is conducted by hand or by computer. The LCF architecture does not allow unknown expressions in goals, so Petersson's Type Theory system does not support backwards proof. Nuprl [6] works explicitly with types only, handling constructions internally.

I have worked out the proofs in this paper by hand. If the type $A$ represents a proposition to be proved, then the initial goal is $p \in A$, for some construction $p$. In proving $p \in A$, each backwards step decomposes the type $A$, incrementally discovering the structure of $p$. The letters $p$, $q$, $r$, ... stand for unknown constructions.

Unification gives a flexible treatment of unknown expressions in goals [19]. An inference rule

$$\frac{\Phi_1(\vec{p}) \quad \cdots \quad \Phi_n(\vec{p})}{\Phi(\vec{p})} \quad ,$$

where $\vec{p}$ stands for a vector of unknowns, specifies a way to reduce a goal $\Phi$ to subgoals $\Phi_1, \ldots, \Phi_n$. Suppose that the goal is $\Psi(\vec{q})$, and that the vector of expressions $\vec{a}$ unifies the goal with the conclusion of the rule: $\Phi(\vec{a})$ is identical to $\Psi(\vec{a})$. Then instantiating the unknowns $\vec{p}$ and $\vec{q}$ reduces $\Psi(\vec{q})$ to $\Phi_1(\vec{a}), \cdots, \Phi_n(\vec{a})$. The unifier $\vec{a}$ may contain new unknowns. Most hand proofs only require trivial unifications.

Backwards proof is natural for both discovering and presenting formal proofs. Consider this backwards proof of the Axiom of Choice. You might compare it with Martin-Löf's forwards proof [14, 15].

It suffices to find some construction *choice* of type

$$\left( \prod_{x \in A} \sum_{y \in B(x)} C(x,y) \right) \to \sum_{f \in \Pi(A,B)} \prod_{x \in A} C(x, f(x)) \ .$$

By product introduction, *choice* could be $\lambda z.p$, where

$$p \in \sum_{f \in \Pi(A,B)} \prod_{x \in A} C(x, f(x)) \ \left[ \ z \in \prod_{x \in A} \sum_{y \in B(x)} C(x,y) \ \right] \ .$$

By sum introduction, $p$ could be the pair $\langle f, q \rangle$, where $f \in \Pi(A,B)$ and $q \in \prod_{x \in A} C(x, f(x))$. First try to find $f$. By product introduction, $f$ could be $\lambda x.r$, where

$$r \in B(x) \ \left[ \begin{array}{l} z \in \prod_{x \in A} \sum_{y \in B(x)} C(x,y) \\ x \in A \end{array} \right] .$$

Product elimination gives $z * x \in \sum_{y \in B(x)} C(x,y)$. By sum elimination, $r$ is $\mathbf{fst}(z * x)$, and $f = \lambda x. \mathbf{fst}(z * x)$; by similar reasoning, $q = \lambda x. \mathbf{snd}(z * x)$. Putting the pieces together gives

$$choice = \lambda z. \langle \lambda x. \mathbf{fst}(z * x), \ \lambda x. \mathbf{snd}(z * x) \rangle \ . \ \blacksquare$$

## 4 Well-founded relations in Type Theory

In Type Theory, the rule of w.f. induction can be stated as

$$\frac{a \in A \qquad s(x, ih) \in P(x) \ \left[ \begin{array}{c} x \in A \\ ih(x', ls) \in P(x') \ [x' \in A; \ ls \in x' \prec x] \end{array} \right] }{\mathbf{wfrec}(s, a) \in P(a)} \ .$$

(Note that the induction hypothesis *ih* is itself hypothetical.) The operator **wfrec** can be used to define functions on $A$. From our understanding of w.f. recursion, it is natural to expect **wfrec** to satisfy the *recursion rule*

$$\frac{\text{(premises as above)}}{\textbf{wfrec}(s,a) = s(a, (x, \textit{ls})\, \textbf{wfrec}(s,x)) \in P(a)} \; .$$

**Definition.** A binary relation $\prec$ on a type $A$ is *well-founded* (w.f.) precisely when the rules of w.f. induction and recursion hold.

A w.f. relation need not be transitive, hence need not be an ordering. It must be irreflexive and asymmetric. The recursion rule has a key advantage: its conclusion does not make use of the propositional variable *ls*. This variable justifies the recursion but plays no role in computation. When using a function $f$ defined by w.f. recursion, the induction rule serves to prove facts about $f$, while the recursion rule allows computation of $f$. By w.f. induction, it is trivial to prove that **wfrec** gives the *unique* solution to this recursion equation.

The symbol **wfrec** denotes a complex Type Theory construction dependent upon the relation $\prec$. A version of **wfrec** is derived for each w.f. relation considered below. For a w.f. relation defined in terms of other w.f. relations, its **wfrec** invokes the **wfrec** of those relations.

How does **wfrec** compare with the fixedpoint operator of domain theory? For $x \in A$, let $[A]_{\prec x}$ denote the subtype of $A$ below $x$, namely $\{x \in A \mid x \prec a\}$. Roughly speaking, w.f. recursion produces a total function of type $A \to B$ given something that maps $x \in A$ to $([A]_{\prec x} \to B) \to B$. Domain theory produces a function of type $A \to B$, given something of type $(A \to B) \to (A \to B)$. So the two recursion operators have similar functionality.

The universe rules of Type Theory allow us to form universes of w.f. relations, with induction and recursion restricted to small types. Abstract over the rules' premises, including implicit ones like $P(x)$ **type**. For the type of a w.f. induction step, use the abbreviation

$$\text{STEP}(A, \prec, P) \equiv \prod_{x \in A} \left( \prod_{x' \in A} x' \prec x \to P * x' \right) \to P * x.$$

The universe $\text{WF}_0$ of small w.f. relations is

$$\sum_{A \in U_0} \sum_{\prec \in A \times A \to U_0} \prod_{P \in A \to U_0} \prod_{\textit{step} \in \text{STEP}(A,\prec,P)} \left\{ \textit{wf} \in \prod_{x \in A} P * x \,\bigg|\, \prod_{x \in A} \textit{wf} * x =_{P*x} \textit{step} * x * (\lambda x'\, \textit{ls}.\textit{wf} * x') \right\}$$

Derivations of w.f. relations can be regarded as constructions within $\text{WF}_0$. The ordering $<$ is an element of $\text{WF}_0$; lexicographic product is a binary operation on $\text{WF}_0$, etc. Members of $WF_0$ have the form $\langle A, \langle \prec, wf \rangle \rangle$. If $P(x) \in U_0$ for any $x \in A$, then w.f. induction and recursion hold under the definition

$$\mathbf{wfrec}(s, a) \equiv wf * \lambda(P) * (\lambda x\, ih.s(x, (x', ls)ih * x' * ls)) * a \ .$$

## 5 Defining Quicksort by well-founded recursion

Before plunging into the formal derivations of w.f. induction and recursion, let us see how to use these rules to reason about recursive functions in Type Theory. Consider an example due to Jan Smith [24]. If $A$ is a type with some total ordering of type $A \times A \to \mathbf{Bool}$, then any list of elements of $A$ can be sorted into ascending order. Quicksort sorts a non-empty list $\mathbf{cons}(a, l)$ by partitioning $l$ into two sublists: one containing the elements that are less or equal to $a$, and one containing the elements that are greater. It recursively sorts these sublists, then concatenates them.

Smith introduces Quicksort within Type Theory by deriving a new principle of primitive recursion. Below, Quicksort is defined by w.f. recursion over a w.f. relation, $\prec$. Each recursive call includes an explicit termination argument involving $\prec$. An unfolding step shows that the resulting function satisfies the usual recursion equations for Quicksort.

Quicksort terminates because the length of the list is smaller in each recursive call. The function *length* is defined as

$$length \equiv \mathbf{listrec}(0, (x, l, u)\,\mathbf{succ}(u)) \ .$$

Thus *length* satisfies the recursion equations

$$\begin{aligned} length(\mathbf{nil}) &= 0 \\ length(\mathbf{cons}(a, l)) &= \mathbf{succ}(length(l)) \ . \end{aligned}$$

Henceforth, functions will be defined by recursion equations whenever these can obviously be translated into a formal definition involving **natrec**, **listrec**, etc. The two equations for *length* give a precise and readable definition.

Define the w.f. relation $\prec$ on lists such that $l' \prec l$ means $length(l') < length(l)$. In other words, $\prec$ is the inverse image of $<$ under the function *length*. Quicksort is defined by recursion over $\prec$. Be careful not to confuse the total ordering on $A$ with the w.f. relation on $\mathbf{List}(A)$.

Partitioning the input list requires a function *filter*. Given a predicate function $pf \in A \to \mathbf{Bool}$, and a list, *filter* returns the list of elements for which $pf$ returns $\mathbf{T}$:

$$\begin{aligned} filter(pf, \mathbf{nil}) &= \mathbf{nil} \\ filter(pf, \mathbf{cons}(a, l)) &= \begin{cases} \mathbf{cons}(a, filter(pf, l)) & \text{if } pf * a = \mathbf{T} \\ filter(pf, l) & \text{if } pf * a = \mathbf{F} \end{cases} \end{aligned}$$

It is straightforward to prove that *filter* does not make the list longer:

$$length(filter(pf, l)) \leq length(l) \ \mathbf{true}$$

A corollary of this, needed to justify the recursion of Quicksort, is (for some construction *qless*)

$$\left[\begin{array}{c} pf \in A \to \mathbf{Bool};\ x \in A;\ l \in \mathbf{List}(A) \\ qless(pf, x, l)\ \in\ length(\mathit{filter}(pf, l)) < length(\mathbf{cons}(a, l)) \end{array}\right],$$

or equivalently

$$qless(pf, x, l)\ \in\ \mathit{filter}(pf, l) \prec \mathbf{cons}(a, l)\ .$$

Two predicate functions provide the ordering used for sorting:

$$\begin{array}{ll} \mathit{before}(a) * b & \text{whether } b \text{ is less than or equal to } a \\ \mathit{after}(a) * b & \text{whether } b \text{ is greater than } a. \end{array}$$

Let $\oplus$ denote the *append* operation on lists (concatenation). We would like to establish the usual recursion equations for Quicksort, namely

$$\begin{aligned} quick(\mathbf{nil}) &= \mathbf{nil} \\ quick(\mathbf{cons}(a, l)) &= (quick(\mathit{filter}(\mathit{before}(a), l))) \\ &\quad \oplus \mathbf{cons}(a,\ quick(\mathit{filter}(\mathit{after}(a), l)))\ . \end{aligned}$$

Defining Quicksort by w.f. recursion requires defining an induction step $s(l, ih)$ in terms of a list $l$ and induction hypothesis

$$ih(l', ls) \in \mathbf{List}(\mathbf{Nat})\ \left[\ l' \in \mathbf{List}(\mathbf{Nat});\ ls \in l' \prec l\ \right]\ .$$

The induction hypothesis allows a recursive call for any list $l'$ smaller than $l$. A proof of $l' \prec l$ must be supplied as an argument. This proof will be a construction, involving *qless*, of type $l' \prec l$. Since the list may be empty or not, the definition of $s$ considers two cases:

$$\begin{aligned} s(\mathbf{nil}, ih) &= \mathbf{nil} \\ s(\mathbf{cons}(a, l), ih) &= ih(\mathit{filter}(\mathit{before}(a), l),\ qless(\mathit{before}(a), a, l)) \\ &\quad \oplus \mathbf{cons}(a,\ ih(\mathit{filter}(\mathit{after}(a), l),\ qless(\mathit{after}(a), a, l))) \end{aligned}$$

Now Quicksort is just $quick \equiv \mathbf{wfrec}(s)$. To produce the familiar recursion equations, unfold *quick* according to the recursion rule for **wfrec**:

$$quick(l) = s(l,\ (l', ls)\, quick(l'))$$

The termination arguments involving *qless* drop out:

$$\begin{aligned} quick(\mathbf{cons}(a, l)) &= s(\mathbf{cons}(a, l),\ (l', ls)\, quick(l')) \\ &= (quick(\mathit{filter}(\mathit{before}(a), l))) \\ &\quad \oplus \mathbf{cons}(a,\ quick(\mathit{filter}(\mathit{after}(a), l))) \end{aligned}$$

Quicksort is correct if its result is always an ordered permutation of its argument. The proof involves w.f. induction, where the induction hypothesis states that *quick* is correct in its recursive calls. Manna and Waldinger verify Quicksort in detail [13].

How efficient is the execution of Quicksort in Type Theory? A detailed analysis of Quicksort's version of **wfrec** is required. Efficiency questions are particularly delicate because Type Theory programs are executed under lazy evaluation. Ideally, Quicksort's recursion equations should be executed directly as rewrite rules: we know that they represent a terminating computation.

## 6 The less-than ordering on the natural numbers

Well-founded induction over the ordering $<$ on natural numbers is the familiar *course-of-values* induction. It is easy to derive from mathematical induction, just as course-of-values recursion is easy to derive from primitive recursion. The relation $<$ is defined to satisfy (remember that $+$ denotes disjoint union!)

$$m < 0 = \bot$$
$$m < \mathbf{succ}(n) = (m =_{\mathbf{Nat}} n \; + \; m < n).$$

The substructure relation for lists and other tree-like types is analogous to $<$. Manna and Waldinger use w.f. induction on this relation for their $l$-expressions [12]. I was surprised to discover that w.f. induction on $l$-expressions has essentially the same derivation as that for $<$. However, we shall see later on that the substructure relation for *any* tree-like type follows from the rules for wellordering types and transitive closure.

### 6.1 Induction

Assume throughout that $P(n)$ is a type for $n \in \mathbf{Nat}$, and assume the induction step

$$step \; \in \; \prod_{n \in \mathbf{Nat}} \left( \prod_{m \in \mathbf{Nat}} m < n \to P(m) \right) \to P(n) \; .$$

To justify w.f. induction, it suffices to find $wf * n \in P(n)$ for $n \in \mathbf{Nat}$. Appealing to the induction step, $wf$ could be $\lambda n. step * n * p(n)$, where

$$p(n) \in \prod_{m \in \mathbf{Nat}} m < n \to P(m) \; .$$

By natural number induction, $p(n)$ could be $\mathbf{natrec}(p_0, p_1, n)$, where

$$p_0 \in \prod_{m \in \mathbf{Nat}} m < 0 \to P(m).$$

and

$$p_1(n, u) \in \prod_{m \in \mathbf{Nat}} m < \mathbf{succ}(n) \to P(m) \; \left[ \begin{array}{l} n \in \mathbf{Nat} \\ u \in \prod_{m \in \mathbf{Nat}} m < n \to P(m) \end{array} \right] \; .$$

Clearly $p_0$ is $\lambda m\, ls.\mathbf{contr}(ls)$. By product introduction and unfolding the less-than relation, $p_1(n, u)$ could be $\lambda m\, ls.p_2$, where

$$p_2 \in P(m) \ \Big[\ n; u;\ m \in \mathbf{Nat};\ ls \in m = n + m < n\ \Big]\ .$$

To save space, I have abbreviated the assumptions $n \in \mathbf{Nat}$ and

$$u \in \prod_{m \in \mathbf{Nat}} m < n \to P(m)$$

as $n$ and $u$. By + elimination, $p_2$ could be $\mathbf{when}(q_1, q_2, ls)$, where

$$q_1(e) \in P(m)\ \Big[\ n; u; m;\ e \in m = n\ \Big]$$

and

$$q_2(ls) \in P(m)\ \Big[\ n; u; m;\ ls \in m < n\ \Big]\ .$$

The induction hypothesis $u$ solves both goals. Replacing $m$ by $n$, the w.f. induction step gives $q_1(e) = step * n * u$; also $q_2(ls) = u * m * ls$. ∎

The definition of $wf$ is too long to work with directly. Instead we can derive a set of equations for it, retaining $p$ as an auxiliary function. Even $p$ is inconveniently complex. It involves the selectors $\mathbf{natrec}$ and $\mathbf{when}$. Therefore it is described by equations where the selectors can be eliminated:

$$\begin{aligned}
p(0) * m * ls &= \mathbf{contr}(ls) & \in P(m) & \quad [ls \in m < 0] \\
p(\mathbf{succ}(n)) * m * \mathbf{inl}(eq) &= step * m * p(n) \in P(m) & & \quad [eq \in m = m] \\
p(\mathbf{succ}(n)) * m * \mathbf{inr}(ls) &= p(n) * m * ls & \in P(m) & \quad [ls \in m < n] \\
wf * n &= step * n * p(n) \ \in P(n)
\end{aligned}$$

Most of derivations of constructions in this paper end with such equations, which summarize the derivation. The following proof of the recursion rule illustrates that the equations provide the needed information in a convenient form.

## 6.2 Recursion

The desired property is

$$wf * n = step * n * (\lambda m\, ls.wf * m)\ \in\ P(n)\ \Big[\ n \in \mathbf{Nat}\ \Big]\ .$$

Unfolding $wf$, it remains to show

$$step * n * p(n) = step * n * (\lambda m\, ls.wf * m)\ \in\ P(n)\ \Big[\ n \in \mathbf{Nat}\ \Big]\ .$$

Canceling (product elimination and introduction), it is enough to show

$$p(n) * m * ls = wf * m\ \in\ P(m)\ \Big[\ n;\ m \in \mathbf{Nat};\ ls \in m < n\ \Big]\ .$$

By equality and product elimination, it is enough to show

$$\prod_{ls \in m < n} p(n) * m * ls =_{P(m)} wf * m \text{ true} \ \Big[ \ n; \ m \in \mathbf{Nat} \ \Big] \ .$$

By natural number induction on $n$, and product introduction, it remains to show

$$p(0) * m * ls = wf * m \ \in \ P(m) \ \Big[ \ m \in \mathbf{Nat}; \ ls \in m < 0 \ \Big] \ ,$$

which holds by the contradiction $m < 0$, and also to show

$$\begin{bmatrix} m \in \mathbf{Nat} \\ n \in \mathbf{Nat} \\ u \in \prod_{ls \in m < n} p(n) * m * ls =_{P(m)} wf * m \\ ls \in m = n \ + \ m < n \\ p(\mathbf{succ}(n)) * m * ls = wf * m \ \in \ P(m) \end{bmatrix} \ .$$

By + elimination, it remains to consider two cases, using the appropriate equation for $p(\mathbf{succ}(n))$:

$$step * n * p(n) = wf * n \ \in \ P(n) \ \Big[ \ m; n; u; \ e \in m = n \ \Big]$$

and

$$p(n) * m * ls = wf * m \ \in \ P(m) \ \Big[ \ m; n; u; \ ls \in m < n \ \Big]$$

The first goal is the equation for $wf$; the second is an instance of the induction hypothesis.

∎

## 7 Subrelations

This paper is less concerned with particular w.f. relations like $<$ as it is with rules for constructing w.f. relations from others. The simplest rule concerns *subrelations* of w.f. relations. Suppose that $\prec$ is a w.f. relation on the type $A$, and that $\ll$ is a subrelation of $\prec$, namely there is a function $f$ satisfying

$$f(x', x, lt) \in x' \prec x \ \Big[ \ x \in A; \ x' \in A; \ lt \in x' \ll x \ \Big] \ .$$

Then $A$ is w.f. by the relation $\ll$.

A typical use of this rule is to justify w.f. induction on the 'properly divides' relation on natural numbers [13]. If $m$ properly divides $n$, then $m < n$. But the main application in Type Theory is when one relation is *logically equivalent* to another. A proposition may be expressed as many different types; this rule provides compatibility between them.

For instance, the proof that $<$ is w.f. uses a particular definition of $<$. You may prefer to work with the logically equivalent definition

$$m \ll n \ \equiv \ \sum_{k \in \mathbf{Nat}} \mathbf{succ}(m \, \mathbf{plus} \, k) = n \ .$$

In order to justify w.f. induction on $\ll$, you need only show that $m \ll n$ implies $m < n$.

## 7.1 Induction

Assume throughout that $P(x)$ is a type for $x \in A$, and assume the induction step

$$step \ \in \ \prod_{x \in A} \left( \prod_{x' \in A} x' \ll x \to P(x') \right) \to P(x).$$

It suffices to find $wf * x \in P(x)$ for $x \in A$. Since the relation $\prec$ is w.f., induction gives $wf = \lambda x.\mathbf{wfrec}(s, x)$, where

$$s(x, ih) \in P(x) \left[ \begin{array}{c} x \in A \\ ih(x', ls) \in P(x') \ [x' \in A; \ ls \in x' \prec x] \end{array} \right].$$

Using the induction step, $s(x, ih)$ could be $step * x * t$, where

$$t \in \prod_{x'} x' \ll x \to P(x').$$

By product introduction, $t$ is $\lambda x'.\lambda lt.ih(x', f(x', x, lt))$, and we have

$$s(x, ih) = step * x * (\lambda x' \ lt.ih(x', f(x', x, lt))). \ \blacksquare$$

## 7.2 Recursion

Since $A$ is w.f. under $\prec$, it satisfies the recursion rule

$$\mathbf{wfrec}(s, a) = s(a, \ (x, ls) \ \mathbf{wfrec}(s, x)).$$

The recursion rule $A$ under $\ll$ is easy to prove, by unfolding $wf$:

$$\begin{aligned} wf * x &= \mathbf{wfrec}(s, x) \\ &= s(x, (x', ls)wf * x') \\ &= step * x * (\lambda x' \ lt.((x', ls)wf * x')(x', f(x', x, lt))) \\ &= step * x * (\lambda x' \ lt.wf * x'). \ \blacksquare \end{aligned}$$

## 8 Inverse image

Suppose that $\prec_B$ is a w.f. relation on the type $B$, and that $f$ is a function such that $f(x) \in B$ if $x \in A$. Define the relation $\prec_A$ on $A$ as the inverse image of $\prec_B$ under $f$:

$$x' \prec_A x \ \equiv \ f(x') \prec_B f(x).$$

A function used for this purpose is often called a rank or measure function. Measure functions play the central role in the induction principle of Boyer and Moore [2], and in my characterization of w.f. types (Section 14).

## 8.1 Induction

Assume throughout that, if $x \in A$, then $P(x)$ is a type and $f(x) \in B$. Also assume

$$step \in \prod_{x \in A} \left( \prod_{x' \in A} x' \prec_A x \to P(x') \right) \to P(x)$$

It suffices to find $wf * x \in P(x)$ for $x \in A$. By reflexivity, equality introduction, substitution, and product elimination, $wf$ could be $\lambda x.wf_2 * \mathbf{eq}$, where

$$wf_2 \in (f(x) =_B f(x)) \to P(x) \ \Big[ \ x \in A \ \Big].$$

Now rename certain occurrences of $x$ as $z$: put

$$R(y) \equiv \prod_{z \in A} (f(z) =_B y) \to P(z) \ .$$

By product elimination, $wf_2$ could be $wf_3 * x$, where

$$wf_3 \in R(f(x)) \ \Big[ \ x \in A \ \Big].$$

W.f. induction on $\prec_B$ gives $wf_3 = \mathbf{wfrec}_B(s, f(x))$, where

$$s(y, ih) \in R(y) \left[ \begin{array}{c} x \in A \\ y \in B \\ ih(y', ls) \in R(y') \ [y' \in B; \ ls \in y' \prec_B y] \end{array} \right].$$

By product introduction, $s(y, ih)$ could be $\lambda z \, e.s_2(z, e)$, where

$$s_2(z, e) \in P(z) \ \Big[ \ x; y; ih; \ z \in A; \ e \in f(z) =_B y \ \Big] \ .$$

The induction step gives $s_2(z, e) = step * z * t(y, ih)$, where

$$t(y, ih) \in \prod_{x' \in A} (f(x') \prec_B f(z)) \to P(x') \ \Big[ \ x; y; ih; z; e \ \Big] \ .$$

Using the assumption $f(z) = y$, product introduction gives $t(y, ih) = \lambda x' \, ls.t_2$, where

$$t_2 \in P(x') \ \Big[ \ x; y; ih; z; \ x' \in A; \ ls \in f(x') \prec_B y \ \Big] \ .$$

By reflexivity and the induction hypothesis, $t_2$ is $ih(f(x'), ls) * x' * \mathbf{eq}$. ∎

Let $p(y) \equiv \mathbf{wfrec}_B(s, y)$. Then we can summarize the derivation with equations:

$$\begin{array}{rll} t(y, ih) &= \lambda x' \, ls.ih(f(x'), ls) * x' * \mathbf{eq} \in \prod_{x' \in A}(f(x') \prec_B y) \to P(x') \\ p(y) * x * e &= step * x * t(y, (y', ls)p(y)) \ \in P(x) & [e \in f(x) =_B y] \\ wf * x &= p(f(x)) * x * \mathbf{eq} & \in \prod_{x \in A} P(x) \end{array}$$

The equation for $p$ uses the recursion rule for $\prec_B$, which is valid under the assumption that $\prec_B$ is w.f. This is used in several summaries below, and will not be pointed out again.

## 8.2 Recursion

The recursion rule for $\prec_A$ follows from the equations above:

$$
\begin{aligned}
wf * x &= p(f(x)) * x * \mathbf{eq} \in P(x) \\
&= step * x * t(f(x), (y', ls)p(y')) \\
&= step * x * (\lambda x'\, ls.p(f(x')) * x' * \mathbf{eq}) \\
&= step * x * (\lambda x'\, ls.wf * x') \quad \blacksquare
\end{aligned}
$$

# 9 Transitive closure

A relation $\prec$ is stronger than its transitive closure $\prec^+$. This means that $\prec^+$ gives a more powerful induction rule, while it is easier to prove that $\prec$ is w.f. We get the best of both worlds by showing that the (irreflexive) transitive closure of any w.f. relation is a w.f. ordering. Suppose that $\prec$ is a w.f. relation on the type $A$. The derivation of induction and recursion works for any definition of the transitive closure $\prec^+$ satisfying

$$
x' \prec^+ x \;\leftrightarrow\; x' \prec x + \left( \sum_{y \in A} x' \prec^+ y \times y \prec x \right) .
$$

The $\to$ direction is particularly important: it means there is a construction *trcases* such that if $x' \in A$, $x \in A$, and $lt \in x' \prec^+ x$, then

$$
trcases(x', x, lt) \;\in\; x' \prec x + \left( \sum_{y \in A} x' \prec^+ y \times y \prec x \right) .
$$

For example, recursion on the natural numbers defines finite powers $\prec^n$ of the relation $\prec$, satisfying

$$
\begin{aligned}
x \prec^0 y &= (x =_A y) \\
x \prec^{\mathbf{succ}(n)} z &= \textstyle\sum_{y \in A} x \prec^n y \times y \prec z .
\end{aligned}
$$

We can now define the transitive closure $\prec^+$, and the reflexive/transitive closure $\prec^*$ (which is obviously not w.f.):

$$
\begin{aligned}
x \prec^+ y &\equiv \textstyle\sum_{n \in \mathbf{Nat}} x \prec^{\mathbf{succ}(n)} y \\
x \prec^* y &\equiv \textstyle\sum_{n \in \mathbf{Nat}} x \prec^n y .
\end{aligned}
$$

The proof for this definition of $\prec^+$ uses the **Nat** elimination rule instead of case analysis on *trcases*.

## 9.1 Induction

Assume that $P(x)$ is a type for $x \in A$. Introduce the abbreviation

$$
Q(x) \equiv \prod_{x' \in A} x' \prec^+ x \to P(x') .
$$

Assume the induction step, which is simply

$$step \in \prod_{x \in A} Q(x) \to P(x) \ .$$

It suffices to find $wf * x \in P(x)$ for $x \in A$. Using the induction step, $wf * x$ could be $step * x * q(x)$, where $q(x) \in Q(x)$. W.f. induction on $x$ gives $q(x) = \mathbf{wfrec}(s, x)$, where

$$\left[ \begin{array}{c} x \in A \\ ih(y, ls) \in Q(y) \ [y \in A; \ ls \in y \prec x] \\ s(x, ih) \in Q(x) \end{array} \right] \ .$$

By product introduction, unfolding $\prec^+$ and $Q$, and sum elimination, $s(x, ih)$ could be $\lambda x' \ lt.s_1$, where

$$s_1 \in P(x') \ \left[ \ x; ih; \ x' \in A; \ lt \in x' \prec^+ x \ \right] \ .$$

Using $trcases$ and $+$ elimination, $s_1$ could be $\mathbf{when}(s_2, s_3, trcases(x', x, lt))$, where

$$s_2(ls) \in P(x') \ \left[ \ x; ih; x'; \ ls \in x' \prec x \ \right]$$

and also

$$s_3(y, lt, ls) \in P(x') \ \left[ \ x; ih; x'; \ y \in A; \ lt \in x' \prec^+ y; \ ls \in y \prec x \ \right] \ .$$

Clearly $s_2(ls) = step * x' * ih(x', ls)$ and $s_3(y, lt, ls) = ih(y, ls) * x' * lt$. ∎

Summarizing,

$$q(x) * x' * lt_0 = \begin{cases} step * x' * q(x') & \text{if } trcases(x', x, lt_0) = \mathbf{inl}(ls) \\ q(y) * x' * lt \ \in \ P(x') & \text{if } trcases(x', x, lt_0) = \mathbf{inr}(\langle y, \langle lt, ls \rangle \rangle) \end{cases}$$

$$wf * x = step * x * q(x) \ \in \ P(x) \ .$$

## 9.2  Recursion

The desired rule is

$$wf * x = step * x * (\lambda x' \ lt.wf * x') \ \in \ P(x) \ \left[ \ x \in A \ \right] \ .$$

Unfolding $wf$ in the left side, and canceling, it suffices to prove

$$\prod_{lt \in x' \prec^+ x} q(x) * x' * lt =_{P(x')} wf * x' \ \left[ \ x \in A; \ x' \in A \ \right] \ .$$

W.f. induction over $\prec$ on $x$, followed by product introduction, gives

$$\left[ \begin{array}{c} x' \in A \\ x \in A \\ ih(y, ls) \in \prod_{lt \in x' \prec^+ y} q(y) * x' * lt =_{P(x')} wf * x' \ [y \in A; \ ls \in y \prec x] \\ lt \in x' \prec^+ x \\ q(x) * x' * lt = wf * x' \ \in \ P(x') \end{array} \right] \ .$$

Using $trcases(x', x, lt)$, it is enough to show both

$$step * x' * q(x') = wf * x' \in P(x') \left[ \; x'; x; ih; \; ls \in x' \prec x \; \right],$$

which follows from the definition of $wf$, and also

$$q(y) * x' * lt = wf * x' \in P(x') \left[ \; x'; x; ih; \; y \in A; \; lt \in x' \prec^+ y; \; ls \in y \prec x \; \right],$$

which follows from the induction hypothesis. ∎

## 10 Disjoint sum

Suppose that $\prec_A$ and $\prec_B$ are w.f. on the types $A$ and $B$ respectively. The relation $\prec$ on the type $A + B$ puts the elements of $A$ before the elements of $B$. With $x', x$ in $A$ and $y', y$ in $B$, it satisfies

$$\begin{aligned} \mathbf{inl}(x') \prec \mathbf{inl}(x) &= x' \prec_A x \\ \mathbf{inl}(x) \prec \mathbf{inr}(y) &= \top \\ \mathbf{inr}(y) \prec \mathbf{inl}(x) &= \bot \\ \mathbf{inr}(y') \prec \mathbf{inr}(y) &= y' \prec_B y \, . \end{aligned}$$

### 10.1 Induction

Assume that $P(z)$ is a type for all $z \in A + B$, and assume the induction step

$$step \in \prod_{z \in A+B} \left( \prod_{z' \in A+B} z' \prec z \to P(z') \right) \to P(z).$$

Before proving $P(z)$ for all $z$ in $A+B$, consider the special case of finding $p(x) \in P(\mathbf{inl}(x))$, for $x \in A$. W.f. induction on $\prec_A$ gives $p(x) = \mathbf{wfrec}_A(p_1, x)$, where

$$p_1(x, ih) \in P(\mathbf{inl}(x)) \left[ \begin{array}{c} x \in A \\ ih_A(x', ls) \in P(\mathbf{inl}(x')) \; [x' \in A; \; ls \in x' \prec_A x] \end{array} \right].$$

The induction step gives $p_1(x, ih_A) = step * \mathbf{inl}(x) * q(x, ih_A)$, where

$$q(x, ih_A) \in \prod_{z' \in A+B} z' \prec \mathbf{inl}(x) \to P(z') \left[ \; x; ih_A \; \right].$$

By product introduction and $+$ elimination, $q(x, ih_A)$ could be $\lambda(\mathbf{when}(p_3, p_4))$, where

$$p_3(x') \in \mathbf{inl}(x') \prec \mathbf{inl}(x) \to P(\mathbf{inl}(x')) \left[ \; x; ih_A; \; x' \in A \; \right]$$

and

$$p_4(y') \in \mathbf{inr}(y') \prec \mathbf{inl}(x) \to P(\mathbf{inr}(y')) \left[ \; x; ih_A; \; y' \in B \; \right].$$

The equations for $\prec$ give $p_3(x') = \lambda ls.ih_A(x', ls)$ and $p_4(y') = \lambda ls.\mathbf{contr}(ls)$. ∎

Now try to find $wf * z \in P(z)$ for $z \in A + B$. By + elimination, $wf$ could be $\lambda(\mathbf{when}(p, r))$, where $p(x)$ is as above and $r(y) \in P(\mathbf{inr}(y))$ for $y \in B$. Reasoning as for $p$, using w.f. induction on $\prec_B$ and then the induction step, gives

$$r(y) = \mathbf{wfrec}_B((y, ih_B) step * \mathbf{inr}(y) * s(y, ih_B), y) ,$$

where

$$\left[ \begin{array}{c} y \in B \\ ih_B(y', ls) \in P(\mathbf{inr}(y')) \ [y' \in B; \ ls \in y' \prec_B y] \\ s(y, ih_B) \in \prod_{z' \in A+B} z' \prec \mathbf{inl}(x) \to P(z') \end{array} \right] .$$

Product introduction followed by case analysis on $z'$, using $p$ again, gives

$$s(y, ih_B) = \lambda(\mathbf{when}((x')\lambda ls.p(x'), (y')\lambda ls.ih_B(y', ls))) . \blacksquare$$

Summarizing,

$$\begin{aligned}
s(y, ih_B) * \mathbf{inl}(x') * ls &= p(x') \in P(\mathbf{inl}(x')) \\
s(y, ih_B) * \mathbf{inr}(y') * ls &= ih_B(y', ls) \in P(\mathbf{inr}(y')) \\
r(y) &= step * \mathbf{inr}(y) * s(y, (y', ls)r(y')) \in P(\mathbf{inr}(y)) \\
q(x, ih_A) * \mathbf{inl}(x') * ls &= ih_A(x', ls) \in P(\mathbf{inl}(x')) \\
q(x, ih_A) * \mathbf{inr}(y') * ls &= \mathbf{contr}(ls) \in P(\mathbf{inr}(y')) \\
p(x) &= step * \mathbf{inl}(x) * q(x, (x', ls)p(x')) \in P(\mathbf{inl}(x)) \\
wf * \mathbf{inl}(x) &= p(x) \in P(\mathbf{inl}(x)) \\
wf * \mathbf{inr}(y) &= r(y) \in P(\mathbf{inr}(y)) .
\end{aligned}$$

## 10.2 Recursion

It suffices to show

$$wf * z = step * z * (\lambda z' \, ls.wf * z') \in P(z) .$$

By + elimination on $z$, it suffices to show

$$wf * \mathbf{inl}(x) = step * \mathbf{inl}(x) * (\lambda z' \, ls.wf * z') \in P(\mathbf{inl}(x)) ,$$

and a similar goal for $\mathbf{inr}(y)$. Unfolding $wf$ and $p$ and canceling, it is enough to show

$$q(x, (x', ls)p(x')) = \lambda z' \, ls.wf * z' \in \prod_{z' \in A+B} z' \prec \mathbf{inl}(x) \to P(z') .$$

Product introduction and + elimination on $z'$, unfolding $q$, gives two goals. We already know the first, $p(x') = wf * \mathbf{inl}(x')$; the second has the contradictory assumption $\mathbf{inr}(y') \prec \mathbf{inl}(x)$.

Proving the goal for $\mathbf{inr}(y)$ is similar; the final case analysis reduces it to the known equations $p(x') = wf * \mathbf{inl}(x')$ and $r(y') = wf * \mathbf{inr}(y')$. $\blacksquare$

# 11 Lexicographic product

The lexicographic product of w.f. relations $\prec_A$ and $\prec_B$ is perhaps the most familiar method of combining w.f. relations. Induction over the product amounts to little more than induction over $\prec_A$ followed by induction over $\prec_B$.

This section actually derives w.f. induction for the sum of a family of types $\sum_{x \in A} B(x)$, of which $A \times B$ is a special case. Suppose that $A$ is a type with a w.f. relation $\prec_A$. Suppose also that $B(x)$ is a family of types indexed by $x \in A$, with a family of w.f. relations $\prec_{B(x)}$. Define the lexicographic relation $\prec$ on the type $\Sigma(A, B)$ as

$$\langle x', y' \rangle \prec \langle x, y \rangle \equiv x' \prec_A x \ + \ (x' =_A x \ \times \ y' \prec_{B(x)} y) \ .$$

## 11.1 Induction

Assume that $P(z)$ is a type for all $z \in \Sigma(A, B)$, and assume the induction step

$$step \in \prod_{z \in \Sigma(A,B)} \left( \prod_{z' \in \Sigma(A,B)} z' \prec z \to P(z') \right) \to P(z).$$

It suffices to find $wf * z \in P(z)$ for $z \in \Sigma(A, B)$. By sum elimination, $wf$ could be $\lambda(\mathbf{split}(wf_1))$, where

$$wf_1(x, y) \in P(\langle x, y \rangle) \ \Big[ \ x \in A; \ y \in B(x) \ \Big] \ .$$

By product elimination, $wf_1(x, y)$ could be $p(x) * y$, where

$$p(x) \in \prod_{y \in B(x)} P(\langle x, y \rangle) \ \Big[ \ x \in A \ \Big].$$

W.f. induction on $\prec_A$ gives $p(x) = \mathbf{wfrec}_A(p_1, x)$, where

$$\left[ \begin{array}{c} x \in A \\ ih_A(x', ls_A) \in \prod_{y \in B(x)} P(\langle x', y \rangle) \ [x' \in A; \ ls_A \in x' \prec x] \\ p_1(x, ih_A) \in \prod_{y \in B(x)} P(\langle x, y \rangle) \end{array} \right] \ .$$

By product introduction, $p_1(x, ih_A)$ could be $\lambda y.q(x, ih_A, y)$, where

$$q(x, ih_A, y) \in P(\langle x, y \rangle) \ \Big[ \ x; \ ih; \ y \in B(x) \ \Big].$$

W.f. induction on $\prec_{B(x)}$ gives $q(x, ih_A, y) = \mathbf{wfrec}_{B(x)}(q_2, y)$, where

$$\left[ \begin{array}{c} x \in A \\ ih_A(x', ls_A) \in \prod_{y \in B(x)} P(\langle x', y \rangle) \ [x' \in A; \ ls_A \in x' \prec x] \\ y \in B(x) \\ ih_B(y', ls_B) \in P(\langle x, y' \rangle) \ [y' \in B(x); \ ls_B \in y' \prec_{B(x)} y] \\ q_2(y, ih_B) \in P(\langle x, y \rangle) \end{array} \right] \ .$$

By the induction step, $q_2(y, ih_B)$ could be $step * \langle x, y \rangle * r(x, ih_A, y, ih_B)$, where

$$r(x, ih_A, y, ih_B) \in \prod_{z' \in \Sigma(A,B)} z' \prec \langle x, y \rangle \to P(z') \left[ \; x; ih_A; y; ih_B \; \right].$$

By product introduction and sum elimination on $z'$, unfolding $\prec$, $r(x, ih_A, y, ih_B)$ could be $\lambda(\mathbf{split}((x', y')\lambda(t)))$, where

$$t \in P(\langle x', y' \rangle) \left[ \begin{array}{l} x; ih_A; y; ih_B \\ x' \in A \\ y' \in B(x') \\ ls \in x' \prec_A x \; + \; (x' =_A x \; \times \; y' \prec_{B(x)} y) \end{array} \right].$$

By + elimination on $ls$, $t$ could be $\mathbf{when}(t_A, \mathbf{split}(t_B), ls)$, where (suppressing the assumptions $x, ih_A, \ldots, y'$)

$$t_A(ls_A) \in P(\langle x', y' \rangle) \left[ \; ls_A \in x' \prec_A x \; \right]$$

and

$$t_B(e, ls_B) \in P(\langle x', y' \rangle) \left[ \; e \in x' =_A x; \; ls_B \in y' \prec_{B(x)} y \; \right]$$

Clearly $t_A(ls_A) = ih_A(x', ls_A) * y'$. For the second goal, replacing $x'$ by $x$ gives $t_B(e, ls_B) = ih_B(y', ls_B)$. ∎

Define $p' \equiv (x', ls)p(x')$. This simplifies the equation about $p$ in the summary:

$$\begin{aligned} r(x, ih_A, y, ih_B) * \langle x', y' \rangle * \mathbf{inl}(ls_A) &= ih_A(x', ls_A) * y' \; \in \; P(\langle x', y' \rangle) \\ r(x, ih_A, y, ih_B) * \langle x', y' \rangle * \mathbf{inr}(e, ls_B) &= ih_B(y', ls_B) \; \in \; P(\langle x', y' \rangle) \\ \\ q(x, ih_A, y) &= step * \langle x, y \rangle * r(x, ih_A, y, (y', ls)q(x, ih_A, y')) \; \in \; P(\langle x, y \rangle) \\ p(x) * y &= q(x, p', y) \; \in \; P(\langle x, y \rangle) \\ wf * \langle x, y \rangle &= p(x) * y \; \in \; P(\langle x, y \rangle) \end{aligned}$$

### 11.2 Recursion

For the recursion rule of $\prec$ on $\Sigma(A, B)$, it suffices to show

$$wf * z = step * z * (\lambda z' \; ls.wf * z') \; \in \; P(z) \left[ \; z \in \Sigma(A, B) \; \right].$$

By sum elimination on $z$, it suffices to show

$$p(x) * y = step * \langle x, y \rangle * (\lambda z' \; ls.wf * z') \; \in \; P(\langle x, y \rangle) \left[ \; x \in A; \; y \in B(x) \; \right].$$

Unfolding $p$, it is enough to show

$$q(x, p', y) = step * \langle x, y \rangle * (\lambda z' \; ls.wf * z') \; \in \; P(\langle x, y \rangle) \left[ \; x \in A; \; y \in B(x) \; \right].$$

Unfolding $q$ and canceling on both sides, it is enough to show

$$r(x, p', y, (y', ls)q(x, p', y')) = \lambda z' \; ls.wf * z' \; \in \prod_{z' \in \Sigma(A,B)} z' \prec \langle x, y \rangle \to P(z') \left[ \begin{array}{l} x \in A; \; y \in B(x) \end{array} \right]$$

By product introduction and sum elimination on $z'$, it is enough to show

$$\begin{bmatrix} x; y;\ x' \in A;\ y' \in B(x');\ ls \in \langle x', y' \rangle \prec \langle x, y \rangle \end{bmatrix}$$
$$r(x, p', y,\ (y', ls)q(x, p', y')) * \langle x', y' \rangle * ls = wf * \langle x', y' \rangle\ \in\ P(\langle x', y' \rangle)$$

By case analysis on $\langle x', y' \rangle \prec \langle x, y \rangle$, unfolding $r$, it remains to show

$$p(x') * y' = wf * \langle x', y' \rangle\ ,$$

which is the equation for $wf$, and

$$q(x, p', y') = wf * \langle x', y' \rangle\ ,$$

which follows from the equations for $wf$ and $p$. ∎

## 12  Lexicographic exponentiation

Suppose that $\prec_A$ is a w.f. relation on $A$. The alphabetic ordering of words in a dictionary can be formalized as the lexicographic relation on lists $[x_1, \ldots, x_n]$, where $x_1, \ldots, x_n \in A$. Define the relation $\triangleleft$ to satisfy $[x_1, \ldots, x_m] \triangleleft [a_1, \ldots, a_n]$ whenever there is some $k$ such that $x_i = a_i$ for $i \leq k$, and either $k = m < n$ or $x_{k+1} \prec_A a_{k+1}$. Note that $\triangleleft$ is not necessarily w.f. — there is a descending chain $[1], [0, 1], [0, 0, 1], \ldots$.

To rule out such chains we must allow only lists whose elements appear in decreasing order. The w.f. relation $\prec$ is simply $\triangleleft$ restricted to the set

$$\left\{ [x_1, \ldots, x_n] \,\middle|\, x_1 \succ_A \cdots \succ_A x_n \right\}\ ,$$

formalized as a sum type of the form $\sum_{l \in \mathbf{List}(A)} D(l)$. I call this the *power type* of $A$, or $\mathbf{Pow}(A)$. Its elements correspond to finite subsets of $A$ whenever $\prec_A$ is a total ordering; furthermore, if $\prec_A$ has order type $\alpha$, then $\prec$ has order type $2^\alpha$.

### 12.1  Definition of power types

The derivation of induction for $\mathbf{Pow}(A)$ is the most complex in this paper. We begin with basic concepts of lists:

$$[x_1, \ldots, x_n] \quad \text{abbreviates} \quad \mathbf{cons}(x_1, \cdots \mathbf{cons}(x_n, \mathbf{nil}) \cdots)\ .$$

The operations of append ($\oplus$) and reverse ($rev$) satisfy by definition

$$\begin{aligned}
\mathbf{nil} \oplus l &= l \\
\mathbf{cons}(x, l_1) \oplus l &= \mathbf{cons}(x, l_1 \oplus l)
\end{aligned}$$

$$\begin{aligned}
rev * \mathbf{nil} &= \mathbf{nil} \\
rev * \mathbf{cons}(x, l) &= (rev * l) \oplus [x]\ .
\end{aligned}$$

The relation $\lhd$ is defined such that

$$l' \lhd \mathbf{nil} = \bot$$
$$\mathbf{nil} \lhd \mathbf{cons}(x,l) = \top$$
$$\mathbf{cons}(x',l') \lhd \mathbf{cons}(x,l) = x' \prec_A x + (x' =_A x \times l' \lhd l) .$$

We have the familiar facts

$$l \oplus \mathbf{nil} = l$$
$$(l_1 \oplus l_2) \oplus l_3 = l_1 \oplus (l_2 \oplus l_3)$$
$$rev * (rev * l) = l$$

The constructions *apls* and *lsap* assert facts about $\lhd$, proved by induction on the lists $l$ and $l'$:

$$apls(l', l'', l) \in l' \oplus l'' \lhd l \rightarrow l' \lhd l$$
$$lsap(l', l, l_2) \in l' \lhd l \oplus l_2 \rightarrow l' \lhd l + \sum_{l_1 \in \mathbf{List}(A)}(l' = l \oplus l_1 \times l_1 \lhd l_2)$$

Normally we build lists by adding elements to the front, using **cons**. Here we also must consider adding elements to the rear. Define

$$\mathbf{rlistrec}(c_0, c_1, l) \equiv \mathbf{listrec}(c_0, (x,l,u)c_1(rev*l, x, u), rev*l) ;$$

it is easy to verify the rule of *reverse list induction*:

$$\frac{l \in \mathbf{List}(A) \quad c_0 \in C(\mathbf{nil}) \quad \begin{array}{c}[l \in \mathbf{List}(A);\ x \in A;\ u \in C(l)]\\ c_1(l,x,u) \in C(l \oplus [x])\end{array}}{\mathbf{rlistrec}(c_0, c_1, l) \in C(l)}$$

Functions defined by reverse list recursion obey the computation rules

$$\mathbf{rlistrec}(c_0, c_1, \mathbf{nil}) = c_0 \in C(\mathbf{nil})$$
$$\mathbf{rlistrec}(c_0, c_1, l \oplus [x]) = c_1(l, x, \mathbf{rlistrec}(c_0, c_1, l)) \in C(l \oplus [x]) .$$

The predicate $D(l)$, defined using **rlistrec**, expresses that the elements of the list $l$ appear in descending order:

$$D(\mathbf{nil}) = \top$$
$$D([x]) = \top$$
$$D(l \oplus [x] \oplus [x']) = x' \prec_A x \times D(l \oplus [x])$$

Evidently $D(l \oplus [x])$ implies $D(l)$; reverse list induction on $l$ gives

$$descap(l_1, l) \in D(l_1 \oplus l) \rightarrow D(l_1) \times D(l)$$
$$endls(l, y, x) \in D(l \oplus [y]) \times l \oplus [y] \lhd [x] \rightarrow y \prec_A^+ x .$$

Recall that $\prec_A^+$ denotes the transitive closure of $\prec_A$.

Define the power type of $A$ as

$$\mathbf{Pow}(A) \equiv \sum_{l \in \mathbf{List}(A)} D(l) .$$

The w.f. relation $\prec$ compares the underlying lists using $\lhd$:

$$\langle l', d' \rangle \prec \langle l, d \rangle = l' \lhd l$$

## 12.2 Induction

Assume that $P(z)$ is a type for $z \in \mathbf{Pow}(A)$; define

$$Q(z) \equiv \prod_{z' \in \mathbf{Pow}(A)} z' \prec z \to P(z') \; ;$$

assume the induction step

$$step \in \prod_{z \in \mathbf{Pow}(A)} Q(z) \to P(z) \; .$$

It suffices to find $wf * z \in P(z)$ for $z \in \mathbf{Pow}(A)$. By the induction step, $wf$ could be $\lambda z.step * z * wf_1$, where $wf_1 \in Q(z)$. By sum elimination on $z$, $wf_1$ could be $\mathbf{split}(wf_2, z)$, where

$$wf_2(l, d) \in Q(\langle l, d \rangle) \; \left[ \; l \in \mathbf{List}(A); \; d \in D(l) \; \right] \; .$$

By product elimination, $wf_2(l, d)$ could be $p(l) * d$, where

$$p(l) \in \prod_{d \in D(l)} Q(\langle l, d \rangle) \; .$$

By reverse list induction, $p(l)$ could be $\mathbf{rlistrec}(p_0, p_1, l)$, where

$$p_0 \in \prod_{d \in D(\mathbf{nil})} Q(\langle \mathbf{nil}, d \rangle) \; ,$$

and

$$p_1(l, x, u) \in \prod_{d \in D(l \oplus [x])} Q(\langle l \oplus [x], d \rangle) \; \begin{bmatrix} l \in \mathbf{List}(A) \\ x \in A \\ u \in \prod_{d \in D(l)} Q(\langle l, d \rangle) \end{bmatrix} \; .$$

In the **nil** goal, unfolding $Q$ gives the contradictory assumption $z' \prec \langle \mathbf{nil}, d \rangle$; thus $p_0$ is $\lambda d\, z'\, ls.\mathbf{contr}(ls)$. By product elimination, $p_1(l, x, u)$ could be $q(x) * l * u$, where

$$q(x) \in \prod_{l \in \mathbf{List}(A)} \left( \prod_{d \in D(l)} Q(\langle l, d \rangle) \to \prod_{d \in D(l \oplus [x])} Q(\langle l \oplus [x], d \rangle) \right) \; \left[ \; x \in A \; \right] \; .$$

Now use w.f. induction over the *transitive closure* of $\prec_A$. Iterating the induction hypothesis proves $Q$ for a descending list of any length, provided its elements are all smaller than $x$. This key step is discussed at length below. Write the recursion operator as $\mathbf{wfrec}_A^+$. Then $q(x)$ could be $\mathbf{wfrec}_A^+(q_1, x)$, where

$$\begin{bmatrix} x \in A \\ ih(x', ls) \in \prod_{l \in \mathbf{List}(A)} \left( \prod_{d \in D(l)} Q(\langle l, d \rangle) \to \prod_{d \in D(l \oplus [x'])} Q(\langle l \oplus [x'], d \rangle) \right) \\ [x' \in A; \; ls \in x' \prec_A^+ x] \\ q_1(x, ih) \in \prod_{l \in \mathbf{List}(A)} \left( \prod_{d \in D(l)} Q(\langle l, d \rangle) \to \prod_{d \in D(l \oplus [x])} Q(\langle l \oplus [x], d \rangle) \right) \end{bmatrix} \; .$$

By product introduction and sum elimination on $z'$, unfolding $Q$ and $\prec$, gives

$$q_1(x, ih) \;=\; \lambda l\, u\, d.\lambda(\mathbf{split}((l', d')\lambda lx.r(ih, l, u, d, x, l', d', lx)))\;,$$

where

$$r(ih, l, u, d, x, l', d', lx) \in P(\langle l', d'\rangle) \left[\begin{array}{l} x; ih;\; l \in \mathbf{List}(A) \\ u \in \prod_{d \in D(l)} Q(\langle l, d\rangle) \\ d \in D(l \oplus [x]) \\ l' \in \mathbf{List}(A) \\ d' \in D(l') \\ lx \in l' \prec l \oplus [x] \end{array}\right].$$

Using *lsap*, case analysis on $l' \prec l \oplus [x]$ gives

$$r(ih, l, u, d, x, l', d', lx) = \mathbf{when}(r_1, \mathbf{split}((l_1)\,\mathbf{split}((e, lx)r_2)), lsap(l', l, [x], lx))$$

where

$$r_1(lx) \in P(\langle l', d'\rangle) \left[\; x; ih; l; u; d; l'; d';\; lx \in l' \prec l \;\right]$$

and, substituting with $l' = l \oplus l_1$,

$$r_2 \in P(\langle l \oplus l_1, d'\rangle) \left[\; x; ih; l; u; d;\; l_1 \in \mathbf{List}(A);\; d' \in D(l \oplus l_1);\; lx \in l_1 \prec [x] \;\right]\;.$$

Since $u$ gives $Q(\langle l, d\rangle)$ for any $d \in D(l)$, the first goal holds with

$$r_1(lx) \;=\; u * \mathbf{fst}(descap(l, [x], d)) * \langle l', d'\rangle * lx\;.$$

(To be absolutely formal I should write $descap(l, [x]) * d$, but $descap(l, [x], d)$ is simpler and clearer. Likewise for *lsap*, *apls*, *endls*.)

In the second goal, the induction step gives $r_2$ as $step * \langle l \oplus l_1, d'\rangle * r_3$, where

$$r_3 \in Q(\langle l \oplus l_1, d'\rangle) \left[\; x; ih; l; u; d; l_1; d'; lx \;\right]\;.$$

Product elimination gives $r_3$ as $s(ih, l, u, d, x, l_1) * lx * d'$, where

$$s(ih, l, u, d, x, l_1) \in l_1 \prec [x] \;\to\; \prod_{d' \in D(l \oplus l_1)} Q(\langle l \oplus l_1, d'\rangle) \left[\; x; ih; l; u; d; l_1 \;\right]\;.$$

Reverse list induction shows that the elements of $l_1$ are all smaller than $x$. Take $s(ih, l, u, d, x, l_1)$ as $\mathbf{rlistrec}(s_0, s_1, l_1)$, where

$$s_0 \in \mathbf{nil} \prec [x] \;\to\; \prod_{d' \in D(l \oplus \mathbf{nil})} Q(\langle l \oplus \mathbf{nil}, d'\rangle) \left[\; x; ih; l; u; d \;\right]$$

and

$$s_1(l, y, v) \in l_1 \oplus [y] \prec [x] \;\to\; \prod_{d' \in D(l \oplus l_1 \oplus [y])} Q(\langle l \oplus l_1 \oplus [y], d'\rangle) \left[\begin{array}{l} x; ih; l; u; d;\; l_1 \in \mathbf{List}(A);\; y \in A \\ v \in l_1 \prec [x] \;\to\; \prod_{d' \in D(l \oplus l_1)} Q(\langle l \oplus l_1, d'\rangle) \end{array}\right].$$

The first goal holds because $l \oplus \mathbf{nil} = l$ and the assumption $u$ gives $Q(\langle l, d \rangle)$; take $s_0 = \lambda lx.u$.

The second goal holds because the w.f. induction hypothesis allows us to append $y$ to the list $l \oplus l_1$, preserving the truth of $Q$. The construction is complex; let us consider it in pieces. Under the assumptions $x, ih, \ldots, v$, recalling that $\oplus$ is associative, note the facts

$$descap(l_1, [y], d') \in D(l) \times D(l_1 \oplus [y])) \; \left[ \; d' \in D(l \oplus (l_1 \oplus [y])) \; \right]$$

$$endls(l, y, x, d_1, lx) \in y \prec_A^+ x \; \left[ \begin{array}{l} d_1 \in D(l_1 \oplus [y]) \\ lx \in l_1 \oplus [y] \triangleleft [x] \end{array} \right]$$

$$apls(l_1, [y], [x], lx) \in l_1 \triangleleft [x] \; \left[ \; lx \in l_1 \oplus [y] \triangleleft [x] \; \right]$$

$$ih(y, ls) * (l \oplus l_1) * (v * lx_1) * d' \; \in \; Q(\langle l \oplus l_1 \oplus [y], d' \rangle) \; \left[ \begin{array}{l} ls \in y \prec_A^+ x \\ lx_1 \in l_1 \triangleleft [x] \\ d' \in D((l \oplus l_1) \oplus [y]) \end{array} \right]$$

The construction using $endls$ to assert $y \prec_A^+ x$ from $l_1 \oplus [y] \triangleleft [x]$ and $D(l_1 \oplus [y])$ is the sole essential appeal to the descending property $D$. These constructions combine to solve the second goal:

$$\begin{aligned} s_1(l_1, y, v) \; = \; & \lambda lx \, d'.ih(y, endls(l, y, x, \mathbf{snd}(descap(l_1, [y], d')), lx)) \\ & * (l \oplus l_1) * (v * apls(l_1, [y], [x], lx)) * d' \; \blacksquare \end{aligned}$$

Define $q' \equiv (x', ls)q(x')$. Summarizing,

$$s(ih, l, u, d, x, \mathbf{nil}) * lx = u \; \in \; \prod_{d \in D(l)} Q(\langle l, d \rangle)$$

$$\begin{aligned} s(ih, l, u, d, x, l_1 \oplus [y]) * lx * d' = \; & ih(y, endls(l, y, x, \mathbf{snd}(descap(l_1, [y], d')), lx)) * \\ & (l \oplus l_1) * (s(ih, l, u, d, x, l_1) * apls(l_1, [y], [x], lx)) * d' \\ & \in Q(\langle l \oplus l_1 \oplus [y], d' \rangle) \end{aligned}$$

$$q(x) * l * u * d * \langle l', d' \rangle * lx_0 = \begin{cases} u * \mathbf{fst}(descap(l, [x], d)) * \langle l', d' \rangle * lx \; \in \; P(\langle l', d' \rangle) \\ \quad \text{if } lsap(l', l, [x], lx_0) = \mathbf{inl}(lx) \quad [lx \in l' \triangleleft l] \\ step * \langle l \oplus l_1, d' \rangle * (s(q', l, u, d, x, l_1) * lx * d') \\ \quad \text{if } lsap(l', l, [x], lx_0) = \mathbf{inr}(\langle l_1, \langle e, lx \rangle \rangle) \\ \quad [l' = l \oplus l_1; \; lx \in l_1 \triangleleft [x]] \end{cases}$$

$$p(\mathbf{nil}) * d * z' * ls = \mathbf{contr}(ls) \; \in \; P(z')$$
$$p(l \oplus [x]) = q(x) * l * p(l) \; \in \; \prod_{d \in D(l \oplus [x])} Q(\langle l \oplus [x], d \rangle)$$
$$wf * \langle l, d \rangle = step * \langle l, d \rangle * (p(l) * d) \; \in \; P(\langle l, d \rangle)$$

The equation for $q(x)$ uses the recursion rule of $\mathbf{wfrec}_A^+$.

## 12.3 Intermezzo

This derivation is related to work in the theory of ordinal recursion. Let $\alpha$ be any ordinal less than $\epsilon_0$. Terlouw derives recursion on $2^\alpha$, using recursion on $\alpha$ to define a functional of a higher type level [27]. To examine the connection with the present derivation of induction over $\prec$ on $\mathbf{Pow}(A)$ from induction over $\prec_A$, our first problem is reconciling the notations. Let $\prec$ (also) denote a wellordering of order type $2^\alpha$, and $\prec_A$ a wellordering of order type $\alpha$, and $[F]_{\prec z}$ the restriction of the function $F$ to arguments below $z$. Let the function $F_0(z)$ defined by the transfinite recursion $F_0(z) = G([F_0]_{\prec z}, z)$. Terlouw expresses $F_0$ using a functional $H$ defined in terms of $G$ by transfinite recursion on $\prec_A$.

Below $x, y, z$ range over ordinals, while $+$ denotes ordinal addition, and $<$ ordinal comparison. Transfinite induction on the proposition $P(x)$ requires that $P$ be *progressive*. The property $\mathrm{Prog}(P)$ is essentially a w.f. induction step: $\mathrm{Prog}(P) \equiv \forall x.\, ((\forall y < x.\, P(y)) \to P(x))$. Terlouw remarks: To prove $P(x)$ for all $x$ up to $\epsilon_0$, let $\alpha$ be arbitrary and prove $\forall x < 2^\alpha.\, P(x)$ using only transfinite induction up to $\alpha$. This follows by proving $\forall x < \alpha.\, B(x)$, where

$$B(x) \equiv \forall z.\, ((\forall y < z.\, P(y)) \to \forall y < z + 2^x.\, P(y))\ .$$

Induction is sound because $\mathrm{Prog}(P)$ implies $\mathrm{Prog}(B)$.

Up to $\epsilon_0$ there is a one-to-one correspondence between ordinals and descending lists of ordinals $2^{\alpha_0} + \cdots + 2^{\alpha_n}$, where $\alpha_0 > \cdots > \alpha_n$. So we can regard any ordinal $z$ uniquely as a list $[\alpha_0, \ldots, \alpha_n]$, and if $x < \alpha_n$, adding $2^x$ to $z$ corresponds to putting $x$ on the end of the list. Thus $B(x)$ is essentially the type of the construction $q(x)$ above. Induction up to the ordinal $\alpha$ proves $B(x)$; w.f. recursion over $\prec_A^+$ establishes $q(x)$.

Terlouw does not incant "propositions as types," but remarks (page 398) that the relationship between the functionals $H$ and $G$ "is nothing other than the functional-analogy" of the relationship between $B$ and $P$. In $H$, the variables $x$ and $z$ are those of $B$, while $F$ denotes a function defined below $z$, of "type" $\forall y < z.\, P(y)$. The construction $q(x) * l * u$ corresponds to $H(x, z, F)$, for $u$ is a function asserting $P$ everywhere below $l$. To clarify the correspondence between $H$ and $q$, define the Type Theory function $h(x, z, F)$ such that $h(x, \langle l, d \rangle, F) \equiv q * l * F * d$. Take the type $A$ to be the natural numbers, and take $x = 0$. Suppressing constructions of type $z' \prec z$, unfold the equations for $q$ and $s$ in the previous subsection. The second equation for $q$ has $l_1 \not\prec [0]$, which is only possible if $l_1 = \mathbf{nil}$ and $l' = l$. This gives

$$h(0, z, F) * z' = \begin{cases} F * (\cdots) * z' & \text{if } z' \prec z \\ \lambda lx.step * z' * (F * \cdots) & \text{if } z' = z \ . \end{cases}$$

Terlouw defines $H$ to satisfy

$$H(0, z, F)(z') = \begin{cases} F(z') & \text{if } z' \prec z \\ G([F]_{\prec z'}, z') & \text{if } z' = z\ . \end{cases}$$

Noting that $[F]_{\prec z}$ corresponds to some construction $F'$ of type $Q(z)$, and that $G([F]_{\prec z}, z)$ corresponds to $step * z * F'$, the two sets of equations are nearly identical. Unfortunately, I cannot carry the correspondence between $H$ and $q$ any further. The rest of the definition of $H$ is more complicated, involving functions for computing ordinal operations in terms of an embedding in the natural numbers. Apparently $H$ builds up its result in a different manner from $q$.

## 12.4 Recursion

The desired property is

$$wf * z = step * z * (\lambda z'\, ls.wf * z') \ \in \ P(z) \ \left[\ z \in A\ \right]\ .$$

Unfolding $wf$ and canceling, it suffices to prove

$$p(l) * d = \lambda z'\, ls.wf * z' \ \in \ Q(\langle l, d\rangle) \ \left[\ l \in \mathbf{List}(A);\ d \in D(l)\ \right]\ .$$

By product elimination, it suffices to show

$$\prod_{d \in D(l)} \prod_{z' \in \mathbf{Pow}(A)} \prod_{ls \in z' \prec \langle l, d\rangle} p(l) * d * z' * ls =_{P(z')} wf * z' \ \mathbf{true}\ \left[\ l \in \mathbf{List}(A)\ \right]\ .$$

By reverse list induction on $l$, followed by product introduction, it is enough to show

$$\mathbf{contr}(ls) = wf * z' \ \in \ P(z') \ \left[\ d;\ z';\ ls \in z' \prec \langle \mathbf{nil}, d\rangle\ \right]$$

(which holds by contradiction), and also

$$\left[\begin{array}{l} l \in \mathbf{List}(A);\ x \in A \\ u \in \prod_{d \in D(l)} \prod_{z' \in \mathbf{Pow}(A)} \prod_{ls \in z' \prec \langle l, d\rangle} p(l) * d * z' * ls =_{P(z')} wf * z' \\ d \in D(l \oplus [x]);\ z' \in \mathbf{Pow}(A);\ ls \in z' \prec \langle l \oplus [x], d\rangle \end{array}\right] \\ q(x) * l * p(l) * d * z' * ls = wf * z' \ \in \ P(z')$$

By sum elimination on $z'$, unfolding $\prec$, it suffices to show

$$\left[\ l; x; u; d;\ l' \in \mathbf{List}(A);\ d' \in D(l');\ lx \in l' \triangleleft l \oplus [x]\ \right] \\ q(x) * l * p(l) * d * \langle l', d'\rangle * lx = wf * \langle l', d'\rangle \ \in \ P(\langle l', d'\rangle)\ .$$

By case analysis with $lsap$ on $l' \triangleleft l \oplus [x]$, unfolding $q$, it suffices to show

$$\left[\ l; x; u; d; l'; d';\ lx \in l' \triangleleft l\ \right] \\ p(l) * \mathbf{fst}(descap(l, [x], d)) * \langle l', d'\rangle * lx = wf * \langle l', d'\rangle \ \in \ P(\langle l', d'\rangle)$$

(which holds by the induction hypothesis), and also

$$\left[\ l; x; u; d;\ l_1 \in \mathbf{List}(A);\ d' \in D(l \oplus l_1);\ lx \in l_1 \triangleleft [x]\ \right] \\ step * \langle l \oplus l_1, d'\rangle * (s(q', l, p(l), d, x, l_1) * lx * d') \\ = wf * \langle l \oplus l_1, d'\rangle \ \in \ P(\langle l \oplus l_1, d'\rangle)\ .$$

By canceling and product elimination, it suffices to show

$$\prod_{lx \in l_1 \prec [x]} \prod_{d' \in D(l \oplus l_1)} s(q', l, p(l), d, x, l_1) * lx * d' =_{Q(\langle l \oplus l_1, d' \rangle)} p(l \oplus l_1) * d' \ \mathbf{true} \quad \begin{bmatrix} l; x; u; d; l_1 \end{bmatrix} \cdot$$

Use reverse list induction on $l_1$, then product introduction, and then unfold $s$. It suffices to show $p(l) = p(l \oplus \mathbf{nil})$, which is trivial, and also

$$\begin{bmatrix} l; x; u; d; \ l_1 \in \mathbf{List}(A); \ y \in A \\ v \in \prod_{lx \in l_1 \prec [x]} \prod_{d' \in D(l \oplus l_1)} s(q', l, p(l), d, x, l_1) * lx * d' =_{Q(\langle l \oplus l_1, d' \rangle)} p(l \oplus l_1) * d' \\ lx \in l_1 \oplus [y] \prec [x]; \ d' \in D(l \oplus l_1 \oplus [y]) \end{bmatrix}$$

$$q(y) * (l \oplus l_1) * (s(q', l, p(l), d, x, l_1) * apls(l_1, [y], [x], lx)) * d'$$
$$= p(l \oplus l_1 \oplus [y]) * d' \ \in \ Q(\langle l \oplus l_1 \oplus [y], d' \rangle)$$

Unfolding $p$ and canceling, it remains to show

$$s(q', l, p(l), d, x, l_1) * apls(l_1, [y], [x], lx) = p(l \oplus l_1) \ \in \ \prod_{d' \in D(l \oplus l_1)} Q(\langle l \oplus l_1, d' \rangle) \ ,$$

which follows by product introduction and the induction hypothesis. ∎

## 13 Wellordering types

If $A$ is a type, and $B(x)$ is a type for each $x \in A$, then there is a type $\mathbf{W}_{x \in A} B(x)$, equivalently $\mathbf{W}(A, B)$. Each value of this *wellordering* type is a tree such that each node is labeled with some element $a$ of $A$, and has a branch for each element $b$ in $B(a)$. A tree has the form $\mathbf{sup}(a, f)$, where $f(y) \in \mathbf{W}_{x \in A} B(x)$ for $y \in B(a)$. In this notation, $a$ labels the node, while $f(b)$ follows the branch for $b$ to a subtree having the same wellordering type.

The natural numbers belong to a simple wellordering type. There are two kinds of node:

- those with no branches represent 0;
- those with one branch represent successor numbers.

A possible definition is

$$\mathbf{Nat} \ = \ \mathbf{W}(\mathbf{Bool}, \mathbf{cond}(\bot, \top)) \ .$$

Lists are a similar wellordering type: the second kind of node contains a member of the list, while the branch points to the rest of the list. Wellordering types can also have infinite branching, as in the second number class of ordinals [15, page 82]. (Note: the w.f. relation defined below is not $<$ on the second number class.)

There is a close connection between wellordering types and well-founded relations, as the terminology suggests. Every wellordering type $W$ is w.f. under the relation $\prec_W$, where $w' \prec_W w$ whenever $w'$ is an immediate subtree of $w$. Formally (by transfinite induction), it is easy to define $\prec_W$ such that

$$w' \prec_W \sup(a, f) \;=\; \{y \in B(a) \mid w' =_W f(y)\} \;.$$

Since the natural numbers can be defined as the wellordering type $N$, with $<$ as the transitive closure of $\prec_N$, w.f. induction for $<$ follows from that of wellordering types. I include the separate proof for $<$ because it is a simple introduction to w.f. relations. There is an apparent circularity: the proof for transitive closure involves **Nat** elimination, the rule for induction over the natural numbers. But this is merely to argue by cases, a number is either 0 or is $\mathbf{succ}(n)$. It does not require w.f. induction for $<$.

## 13.1 Induction

Assume that $P(w)$ is a type for $w \in \mathbf{W}(A, B)$, and assume the induction step

$$step \in \prod_{w \in \mathbf{W}(A,B)} \left( \prod_{w' \in \mathbf{W}(A,B)} w' \prec w \to P(w') \right) \to P(w).$$

To justify w.f. induction, it suffices to find $wf * w \in P(w)$ for $w \in \mathbf{W}(A, B)$. By transfinite induction, $wf$ could be $\lambda w.\mathbf{transrec}(p, w)$, where

$$p(a, f, u) \in P(\sup(a, f)) \left[ \begin{array}{ll} a \in A & \\ f(y) \in \mathbf{W}(A, B) & [y \in B(a)] \\ u(y) \in P(f(y)) & [y \in B(a)] \end{array} \right]$$

Using the induction step, $p(a, f, u)$ could be $step * \sup(a, f) * p_1$, where

$$p_1 \in \prod_{w' \in \mathbf{W}(A,B)} w' \prec \sup(a, f) \to P(w') \; \left[ \; a; f; u \; \right].$$

Unfolding $\prec$, product introduction gives $p_1 = \lambda w' \, ls.p_2(ls)$, where

$$p_2(ls) \in P(w') \; \left[ \; a; f; u; \; w' \in \mathbf{W}(A,B); \; ls \in \{y \in B(a) \mid w' = f(y)\} \; \right].$$

Using subset elimination and replacing $w'$ by $f(y)$, the goal becomes

$$p_2(y) \in P(f(y)) \; \left[ \; a; f; u; \; w' \in \mathbf{W}(A,B); \; y \in B(a) \; \right].$$

Appealing to the induction hypothesis, $p_2(y)$ is $u(y)$. ∎

Summarizing,

$$wf * \sup(a, f) = step * \sup(a, f) * (\lambda w' \, ls.wf * f(ls)) \;\in\; P(\sup(a, f)) \;.$$

## 13.2 Recursion

The desired property is

$$wf * w = step * w * (\lambda w'\, ls.wf * w') \in P(w) \,.$$

By W-elimination, $w$ must have the form $\mathbf{sup}(a, f)$, in

$$wf * \mathbf{sup}(a, f) = step * \mathbf{sup}(a, f) * (\lambda w'\, ls.wf * w') \in P(\mathbf{sup}(a, f)) \left[ \begin{array}{c} a \in A \\ f(y) \in \mathbf{W}(A, B) \ [y \in B(a)] \end{array} \right] \,.$$

Unfolding the left side, canceling, and unfolding $\prec$, it remains to show

$$wf * f(ls) = wf * w' \in P(w') \left[ \begin{array}{c} a; f \\ w' \in \mathbf{W}(A, B) \\ ls \in \{y \in B(a) \mid w' = f(y)\} \end{array} \right] \,.$$

By subset elimination, it remains to show

$$wf * f(y) = wf * w' \in P(w') \left[\, a; f; w';\ y \in B(a);\ w' = f(y)\ \mathbf{true}\,\right] \,.$$

This holds because $w'$ equals $f(y)$. ∎

# 14 A characterization of well-founded relations

Classically, a relation $\prec$ on a set $A$ is w.f. whenever there are no infinite descending chains $a_0 \succ a_1 \succ a_2 \cdots$. Another classical characterization is that every nonempty subset of $A$ contains a minimal element [13]. I know of no constructive derivation of w.f. induction from either of these characterizations. However, there is a constructive characterization:

**Theorem.** *A relation $\prec_A$ on a type $A$ is w.f. if and only if there is some wellordering type $W$, and rank function $wof \in A \to W$, such that $\prec_A$ is logically equivalent to the inverse image of $\prec_W$.*

**Proof.** The *if* part is immediate, because every wellordering type is w.f., and taking an inverse image preserves this property. To show the converse requires constructing a wellordering type and rank function whose inverse image is $\prec_A$. Each node of the wellordering type is labeled by an element of $A$ and has one branch for every predecessor $a'$ of $a$: define

$$W \equiv \mathop{\mathbf{W}}_{a \in A} \left( \sum_{x \in A} x \prec_A a \right) \,.$$

The rank function $wof$ is defined by w.f. recursion, such that $wof * a$ is a tree rooted in $a$ and with branches to subtrees $wof * a'$ for every predecessor $a'$. There is a branch for each pair $\langle a', ls \rangle$ where $a' \in A$ and $ls \in a' \prec_A a$, so the definition involves **split**:

$$\begin{aligned} s(a, ih) &\equiv \mathbf{sup}(a, \mathbf{split}(ih)) \\ wof &\equiv \lambda(\mathbf{wfrec}(s)) \,. \end{aligned}$$

Unfolding *wof* using the recursion rule gives

$$\begin{aligned} wof * a &= \mathbf{wfrec}(s, a) \\ &= s(a, (x, ls)\,\mathbf{wfrec}(s, x)) \\ &= s(a, (x, ls)\,wof * x) \\ &= \mathbf{sup}(a, \mathbf{split}((x, ls)\,wof * x)) \ . \end{aligned}$$

By a trivial transfinite recursion, the function $aof \in W \to A$ maps any element $\mathbf{sup}(a, f)$ to $a$: put $aof \equiv \lambda(\mathbf{transrec}((a, f, e)a)$. Then *wof* is one-to-one, since $aof * (wof * a) = a$ for all $a$. Also

$$\begin{aligned} wof * a' \prec_W wof * a &= wof * a' \prec_W \mathbf{sup}(a, \mathbf{split}((x, ls)\,wof * x)) \\ &= \left\{ y \in \sum_{x \in A} x \prec_A a \;\middle|\; wof * a' =_W \mathbf{split}((x, ls)\,wof * x, y) \right\} \end{aligned}$$

To show that $\prec_A$ is logically equivalent to the inverse image under *wof* of $\prec_W$, it suffices to show

$$a' \prec_A a \;\leftrightarrow\; wof * a' \prec_W wof * a \;\mathbf{true} \;\begin{bmatrix} a \in A;\; a' \in A \end{bmatrix} \ .$$

For the $\to$ direction, it is enough to show

$$\left\{ y \in \sum_{x \in A} x \prec_A a \;\middle|\; wof * a' =_W \mathbf{split}((x, ls)\,wof * x, y) \right\} \mathbf{true} \;\begin{bmatrix} ls \in a' \prec_A a \end{bmatrix}$$

which is immediate, putting $\langle a', ls \rangle$ for $y$.

For the $\leftarrow$ direction, by product introduction, subset elimination, and sum elimination, it is enough to show

$$a' \prec_A a \;\mathbf{true} \;\begin{bmatrix} x \in A \\ ls \in x \prec_A a \\ e \in wof * a' =_W wof * x \end{bmatrix} \ .$$

Since *wof* is one-to-one, $a' = x$. ∎

The wellordering types generalize the notion of ordinal. The theorem generalizes the notion of order type: each w.f. relation has a normal form, the inverse image of a wellordering type. This characterization of w.f. relations seems less natural than the classical ones.

## 15  Well-founded relations in the literature

People have proved program termination using a variety of w.f. relations, most of which are easily constructed in this Type Theory framework. The lexicographic product of w.f. relations appears often, but sometimes in disguise. In deriving a unification algorithm, Manna and Waldinger [12] define the "unification ordering" $\prec_{\mathrm{un}}$ on pairs of expressions. Let $vars(x)$ denote the (finite) set of variables in an expression, and $\prec$ the (w.f.) substructure relation on expressions. Manna and Waldinger define $\langle x', y' \rangle \prec_{\mathrm{un}} \langle x, y \rangle$ to hold whenever

$$vars(x') \cup vars(y') \subset vars(x) \cup vars(y)$$

or
$$vars(x') \cup vars(y') = vars(x) \cup vars(y) \quad \text{and} \quad x' \prec x \ .$$

The proper subset relation is w.f. on finite sets: if $|a|$ denotes the cardinality of $a$, then $a \subset b$ implies $|a| < |b|$. So $\prec_{\text{un}}$ is also w.f.: it is a subrelation of the inverse image of a lexicographic product, mapping each $\langle x, y \rangle$ to the lexicographically ordered pair $\langle |vars(x) \cup vars(y)|, x \rangle$.

Generalizing the lexicographic product from $A \times B$ to the sum $\sum_{x \in A} B(x)$ produces many more w.f. relations. The *second number class* is the set of all countable ordinals, and is the smallest class of ordinals containing 0 and closed under successor and countable limits [4]. These three principles underlie standard systems of ordinal notations such as Kleene's $O$ [23]. In Type Theory, take $\bot$ for 0, and $A + \top$ for successor, and $\sum_{n \in \mathbf{Nat}} A(n)$ for limit. Then any notation for a constructive ordinal can be translated into a w.f. relation of the same order type.

A concrete application of lexicographic product for both $\Sigma$ and $\times$ is the *stepped lexicographic ordering* [7, page 474] on $n$-tuples of $A$:

$$A^\omega \equiv \sum_{n \in \mathbf{Nat}} A^n \qquad \text{where} \qquad \begin{aligned} A^0 &= \top \\ A^{\mathbf{succ}(n)} &= A \times A^n \end{aligned}$$

The empty relation on the type $\top$ is trivially w.f., so the relation $\prec_{A^n}$ on the finite power $A^n$ is w.f. by induction on $n$. Each element of $A^\omega$ is an $n$-tuple, represented in the form $\langle n, \langle x_1, \ldots, \langle x_n, 0 \rangle \rangle \rangle$. Under the w.f. relation on $A^\omega$, shorter tuples preceed longer ones and tuples of equal length are compared lexicographically.

Now consider lexicographic exponentiation of the relation $\prec_A$. Recall that elements of the power type $\mathbf{Pow}(A)$ are lists $[x_1, \ldots, x_n]$ with elements in descending order $x_1 \succ_A \cdots \succ_A x_n$. The w.f. relation on power types differs from that on $A^\omega$, for $[1, 0] \prec [2]$ even though $[1, 0]$ is the longer list.

A two-argument form of exponentiation is sometimes seen [26]. Suppose $B$ is a type with w.f. relation $\prec_B$. An element of the type $B^A$ is a list of pairs from $A \times B$, descending under $\prec_A$:

$$\left\{ [\langle x_1, y_1 \rangle, \ldots, \langle x_n, y_n \rangle] \,\middle|\, x_1 \succ_A \cdots \succ_A x_n \right\} \qquad \begin{aligned} x_1, \ldots, x_n &\in A \\ y_1, \ldots, y_n &\in B \end{aligned}$$

This list represents a finite function from $A$ to $B$. The w.f. relation on lists is that for $\mathbf{Pow}(A \times B)$, comparing each pair $\langle x_i, y_i \rangle$ under the relation for $A \times B$. Clearly $x_1 \succ_A \cdots \succ_A x_n$ implies $\langle x_1, y_1 \rangle \succ_{A \times B} \cdots \succ_{A \times B} \langle x_n, y_n \rangle$; formalizing this implication as a Type Theory function defines the w.f. relation for $B^A$ as an inverse image of that for $\mathbf{Pow}(A \times B)$. In fact we could generalize exponentiation to the "finitary product" of the family of types $B(x)$ for $x \in A$, as an inverse image of $\mathbf{Pow}(\Sigma(A, B))$. If $\prec_A$ has order type $\alpha$ and $\prec_B$ has order type $\beta$ then the relation on $B^A$ has order type $\beta^\alpha$.

If $\prec_A$ is a total ordering, then taking $B$ as the natural numbers gives Dershowitz and Manna's [7] ordering on finite multisets of $A$; define $\mathbf{M}(A) \equiv \mathbf{Nat}^A$. Informally, a *multiset* is like a set except that multiple occurrences of an element are significant. Suppose that $a_1 \prec_A \cdots \prec_A a_n$, where $a_i \in A$ for $1 \leq i \leq n$. Then the finite multiset containing $m_i$ occurrences of $a_i$ is represented in $\mathbf{M}(A)$ as $[\langle a_1, m_1 \rangle \ldots, \langle a_n, m_n \rangle]$.

The type $\mathbf{M}^*(A)$ of *nested multisets* over $A$ contains finite multisets with members drawn from both $A$ and $\mathbf{M}^*(A)$. Nesting is finite: $\mathbf{M}^*(A)$ is built up in stages using the family $\mathbf{M}^n(A)$ of multisets nested to depth $n$. The w.f. derivations for $+$ and $\Sigma$ give the appropriate w.f. relation for $\mathbf{M}^*(A)$. Define

$$\mathbf{M}^*(A) \equiv \sum_{n \in \mathbf{Nat}} \mathbf{M}^n(A) \qquad \text{where} \qquad \begin{aligned} \mathbf{M}^0(A) &= A \\ \mathbf{M}^{\mathbf{succ}(n)} &= \mathbf{M}^n(A) + \mathbf{M}(\mathbf{M}^n(A)) \end{aligned}$$

The definition of $\mathbf{M}^*(A)$ will be complete once we have defined operations like $\cup$ and $\{x\}$ for nested multisets, expressing each multiset as $\langle n, \mathbf{M}^n(A) \rangle$ for the smallest $n$ possible.

The type $\mathbf{M}^*(\top)$ is essentially Gentzen's notation [9] for the ordinals up to $\epsilon_0$. While order type $\epsilon_0$ is larger than necessary for most termination questions in computer science, Dershowitz and Manna give several examples where multisets allow simpler proofs. Often the only alternative ordering is the inverse image of $<$ under a complex and subtle arithmetic function [2].

As defined here, the type $\mathbf{M}(A)$ contains only those multisets whose elements are pairwise related under $\prec_A$. If $\prec_A$ is not a total ordering, then some multisets are excluded. I doubt that this restriction will seriously complicate program proofs. The typical partial ordering is an inverse image of $<$. For example, multisets are often used with the subtree ordering on trees; a termination proof using multisets of trees is easily changed to one using multisets of heights of trees. No problems arise in the examples by Dershowitz and Manna [7]. Still, I would like to eliminate the restriction by proving, in Type Theory, that the multiset ordering is w.f. Their proof is far from constructive — it appeals to König's Infinity Lemma — but perhaps can be formalized using wellordering types.

Boyer and Moore present a particularly interesting termination question [2, pages 67–71]. Calling the recursive function $norm(x)$ puts the expression $x$ into a certain normal form. Its termination is not obvious because it sometimes calls itself with a larger expression than it was called with. Boyer and Moore justify the definition of *norm* using a complicated measure function. The LCF system, based on domain theory, can deal with partial functions; it allows a different termination proof, using a lemma that $norm(x)$ terminates for certain $x$. I have proved the termination of *norm* in Type Theory by defining a relation corresponding to *norm*'s recursive calls, and deriving w.f. induction and recursion for this relation [20]. The Type Theory proof bears a striking resemblance, in its outer structure, to the LCF proof; furthermore, it suggests an obviously total alternative definition of *norm*.

# 16 Questions

Type Theory seems ideal for working with w.f. recursion and induction. The rules and constructions make computational sense, having as natural a feel as can be expected from a formal notation. I am generally satisfied with this theory of w.f. relations, and look forward to using it in computer proofs. However, some questions present themselves.

## 16.1 Computation on proof objects

The ability to represent propositions as types allows for a compact logical system, since type operators serve also as logical connectives. Its drawback is that uninteresting propositional constructions can complicate computational ones. Treating the proposition $x' \prec x$ as a type causes complications, even though the recursion rule allows $\mathbf{wfrec}(s, a)$ to be computed without computation on elements of $x' \prec x$. Recall the rule of w.f. induction:

$$\cfrac{a \in A \qquad \left[\begin{array}{c} x \in A \\ ih(x', ls) \in P(x') \ [x' \in A;\ ls \in x' \prec x] \\ s(x, ih) \in P(x) \end{array}\right]}{\mathbf{wfrec}(s, a) \in P(a)}$$

Completely eliminating the dependence of $ih$ on $ls$ would give the rule

$$\cfrac{a \in A \qquad \left[\begin{array}{c} x \in A \\ ih(x') \in P(x') \ [x' \in A;\ x' \prec x\ \mathbf{true}] \\ s(x, ih) \in P(x) \end{array}\right]}{\mathbf{wfrec}(s, a) \in P(a)}$$

The conclusion of the recursion rule, which is now

$$\mathbf{wfrec}(s, a) = s(a, (x, ls)\,\mathbf{wfrec}(s, x)) \in P(a),$$

would simplify to

$$\mathbf{wfrec}(s, a) = s(a, \mathbf{wfrec}(s)) \in P(a).$$

For the characterization of w.f. relations, the type $W$ would become

$$W \equiv \mathop{\mathbf{W}}_{a \in A} \{x \in A \mid x \prec_A a\}\ ,$$

and the rank function $wof$ would be just $\lambda(\mathbf{wfrec}(\mathbf{sup}))$.

To achieve all this requires deriving w.f. induction from an induction step that has the slightly stronger induction hypothesis

$$ih(x') \in P(x') \left[\ x' \in A;\ x' \prec x\ \mathbf{true}\ \right]\ .$$

The proof of $ih(x')$ may no longer use a particular element $ls$ of $x' \prec x$, though assuming its truth. I doubt that this can be done in general. Though the derivations for subrelations and inverse images use *ls* only for appealing to w.f. induction on another relation, most other derivations depend crucially on *ls*.

Consider *decidable* relations defined as $x' \prec x \equiv (f(x',x) = \mathbf{T})$, where $f$ is a computable function. The rule of equality elimination prevents any dependence on an element of $x' \prec x$. Many w.f. relations are decidable: for every constructive ordinal $\alpha$, there is a recursive relation corresponding to $<$ for the ordinals below $\alpha$ [23]. But it is unpleasant to impose decidability when none of the derivations appeal to it; furthermore, the subtree relation on infinitely branching wellordering types is clearly undecidable.

Dybjer suggests a radical alternative [8]: reason about programs in an untyped Logical Theory of propositions. Type Theory interprets propositions as types; the Logical Theory takes propositions as primitive and adopts the more conventional interpretation of types as predicates [25]. Because there are no type constraints, the fixedpoint combinator $Y$ exists and can define recursive functions. Proving that a function has a type amounts to proving its termination. Well-founded induction is still required, but $Y$ takes the place of well-founded recursion.

### 16.2  Does induction entail recursion?

Each derivation of a w.f. relation consists of a derivation of w.f. induction, followed by a proof of the recursion rule for the resulting construction. In every case I have encountered, the proof of the recursion rule closely follows the derivation of induction. It is natural to imagine that the recursion rule always follows from induction, yet this seems impossible to prove. Nor is there an obvious counterexample.

### Acknowledgements

Due to the new ideas involved in Type Theory, and the scarcity of published material, I have relied more than usual on discussions with other people. Particular thanks are due to Peter Aczel and Robert Constable, who visited Cambridge, answered many questions, and provided literature. Martin Hyland gave important advice about ordinals. Thanks also to Michael Fourman, Per Martin-Löf, Bengt Nordström, Kent Petersson, Dana Scott, Jan Smith, Richard Waldinger, and Glynn Winskel.